\documentclass[greek,english,a4paper,fleqn,titlepage,DIV12]{scrartcl}

\usepackage[dvips]{changebar}
\usepackage{graphicdef} 
\usepackage{def}
\numberwithin{equation}{section} 
\usepackage[english]{babel}
\usepackage{cite} 
\usepackage{enumerate} 
\usepackage{url}
\usepackage{ae}
\usepackage{fancyhdr}

\renewcommand{\abstract}[1]{\publishers{%
    \begin{minipage}[h]{\textwidth}%
       \normalfont \normalsize {#1}%
    \end{minipage}}}

\makeatletter
\def\rcsInfo $Id#1$ {}
\makeatother
\rcsInfo $Id: main.tex,v 3.2 2001/06/26 09:57:15 guest Exp guest $

\titlehead{ULM-TP/01-5\\June 2001}
\title{\Large{Parabolic maps with spin:}\\[1ex]
  \Large{Generic spectral statistics with non-mixing classical limit}}
\author{
  \large{Grischa Haag\footnote{E-mail address: {\tt
        grischa.haag@physik.uni-ulm.de}} \ and \ Stefan
    Keppeler\footnote{E-mail address: {\tt
        stefan.keppeler@physik.uni-ulm.de}}}}
\date{\normalsize{Abteilung Theoretische Physik\\
    Universit\"at Ulm, Albert-Einstein-Allee 11\\
    D-89069 Ulm, Germany}}

\begin{document}
\abstract{We investigate quantised maps of the torus whose classical
  analogues are ergodic but not
  mixing. Their quantum spectral statistics shows non-generic
  behaviour, \ie it does not follow random matrix theory (RMT). By
  coupling the map to a spin~${\nicefrac{1}{2}}$, which
  corresponds to changing the quantisation without altering the
  classical limit of the dynamics on the torus, we numerically observe
  a transition to RMT statistics. The results are interpreted in terms
  of semiclassical trace formulae for the maps with and without spin
  respectively. We thus have constructed quantum systems with
  non-mixing classical limit which show generic (\ie RMT) spectral
  statistics. We also discuss the analogous situation for an almost
  integrable map, where we compare to Semi-Poissonian statistics.}

\maketitle

\rcsInfo  $Id: intro.tex,v 3.1 2001/06/26 09:56:48 guest Exp guest $
\section{Introduction}
\label{sec:introduction}

A long standing problem in quantum chaology is the precise formulation
of the conjecture of Bohigas, Giannoni and Schmit
(BGS)~\cite{BohGiaSch84} which states that, generically, the quantum
spectral statistics of systems whose classical analogue is chaotic can
be described by the average statistical behaviour of the eigenvalues
of large random matrices.
In contrast, Berry and Tabor~\cite{BerTab77b} conjectured
that for integrable systems the statistical distribution of
eigenvalues is that of a Poisson process.

Whereas in the latter case integrability is precisely defined by the
Liouville-Arnold theorem~\cite{Arn78},
there is no general consensus as to what should be sufficient
conditions for the BGS conjecture to hold. It is generally believed
that hyperbolicity, \ie positive Lyapunov exponents, is sufficient for
observing statistics as in random matrix theory (RMT). 
However, since this is a very strong
property, one might ask whether weaker conditions would suffice.
Actually, this question has already been raised by Bohigas, Giannoni and
Schmit~\cite{BohGiaSch84}, who originally formulated their conjecture
for K-systems.  The weakest possible property in the hierarchy of
chaos is ergodicity which, roughly speaking, means that almost
everywhere in phase space
time averages are equal to a phase space average.  It is
sometimes argued that ergodicity is too weak for implying RMT
statistics but that instead one should also require the decay of
correlations, \ie mixing or even the K-property.

In order to shed some light on this problem we investigate parabolic
maps of the two-torus, namely Kronecker and skew maps, which are
(uniquely) ergodic but satisfy no stronger condition of chaoticity. In
particular, they are not mixing, \ie in general correlations do not
decay, and have zero topological entropy and hence zero Kolmogorov-Sinai
entropy. Thus, these maps serve as good examples for soft chaos.

The quantised versions of these maps are known to show non-generic
spectral statistics. Even worse, it has been proven~\cite{BacHaa99}
that some statistical functions do not exist in the semiclassical limit.
Although this might be considered as evidence for ergodicity not being
a sufficient condition for RMT statistics, we argue that the
non-generic behaviour is due to degeneracies in both the quantum
spectrum and the spectrum of classical periodic orbits. These
degeneracies can be lifted without changing the characteristic
properties of the classical system by perturbing the quantum system in
higher semiclassical order, namely by coupling the quantum map to a
spin~$\nicefrac{1}{2}$. We show that in this way one can construct a
family of quantum maps which all have a non-mixing classical limit but
numerically show RMT statistics. We analyse the lifting of
degeneracies both numerically and in terms of semiclassical trace
formulae.  Thus we provide evidence that generically (which should at
least imply the absence of systematic degeneracies both quantum
mechanically and semiclassically) ergodicity might be a strong enough
condition for implying the BGS conjecture.

For skew maps we also investigate an almost
integrable situation. The level
spacing distribution for the almost integrable skew map with spin is
close to Semi-Poissonian behaviour. This numerical finding is supported 
by a semi-heuristic analytical calculation of the value of the form 
factor $K(\tau)$ at $\tau=0$.

The paper is organised as follows. First we introduce classical and
quantised Kronecker maps (section~\ref{sec:kronecker-map}
and~\ref{sec:quant-mech-kron} respectively), which are irrational
translations on the two-torus and thus represent the simplest possible
ergodic system to be studied classically and quantum mechanically. In
section~\ref{sec:trace-formula} we discuss their non-generic spectral
statistics and relate it to degeneracies of classical periodic orbits
using a trace formula. The following sections are devoted to the study
of Kronecker maps with spin, which show generic spectral
statistics (section~\ref{sec:maps-with-spin}). We also discuss how a
higher order quantum perturbation lifts the degeneracy of periodic
orbits in a semiclassical trace formula (section
\ref{sec:trace-formula-with-spin}). Then in section~\ref{sec:skew-map}
we turn our attention to the slightly more complicated skew maps. Quantum 
skew maps also show non-generic spectral statistics.  But
again we observe a change to RMT statistics if we add a spin
contribution. Almost integrable skew maps are discussed in 
section~\ref{sec:skew-map-pseudo}. We conclude with some remarks in
section~\ref{sec:conclusions}.  The proof of the Egorov property for
the Kronecker map and the trace formula for the skew map are discussed
in two appendices.

 
\rcsInfo  $Id: kronecker.tex,v 3.1 2001/06/26 10:02:19 guest Exp guest $
\section{Kronecker maps}
\label{sec:kronecker-map}
The simplest example for parabolic maps are 
translations on the two-torus given by
\begin{equation}
  \label{eq:kronecker-map}
  \begin{array}{ccccc}
    \Kron:&\TT&\longrightarrow&\TT& \\[1ex]
    &
    \begin{pmatrix}
      p \\
      q \\
    \end{pmatrix}
    &\longmapsto&
    \begin{pmatrix}
      p+\beta \\
      q+\alpha \\
    \end{pmatrix}
    &\pmod 1 \ ,
  \end{array}
\end{equation}
with $\alpha,\beta\in\Sone$. If $1,\alpha$ and $\beta$ are linearly
independent over $\Z$ we have irrational translations on the two-torus
which are
also called Kronecker maps. In these cases we get uniquely ergodic,
but non-mixing maps (see, \eg, 
\cite{cornfeldfominsinai,katokhasselblatt}). 
For $\alpha,\beta\in\Q$ the maps are periodic. In all other cases the torus
splits into families of invariant curves.

For later reference it will be important to study the symmetries of
$\Kron$. There are several symplectic symmetries $\pi$
fulfilling
\begin{equation}
  \label{eq:symplectic-symmetries}
  \pi\circ \Kron\circ \pi^{-1}=\Kron
  \ .
\end{equation}
First of all we have invariance under inversion
\begin{equation}
  \label{eq:k-inversion}
  I_{\phantom{q}}:\phantom{(\gamma)}
  \begin{pmatrix}
    p \\
    q \\
  \end{pmatrix}
  \longmapsto
  \begin{pmatrix}
    -p \\
    -q \\
  \end{pmatrix}
  \ \pmod 1
  \ .
  \phantom{\ ,\ \gamma\in\Sone}
\end{equation}
Furthermore, there are two continuous families of symmetry operations: 
the maps commute with both translations in position
\begin{equation}
  \label{eq:k-position-symmetry}
  T_q(\gamma):
  \begin{pmatrix}
    p \\
    q \\
  \end{pmatrix}
  \longmapsto
  \begin{pmatrix}
    p \\
    q+\gamma \\
  \end{pmatrix}
  \ \pmod 1
  \ ,\quad \gamma\in\Sone
  \phantom{\ .}
\end{equation}
and translations in momentum
\begin{equation}
  \label{eq:k-momentum-symmetry}
  T_p(\delta):
  \begin{pmatrix}
    p \\
    q \\
  \end{pmatrix}
  \longmapsto
  \begin{pmatrix}
    p+\delta \\
    q \\
  \end{pmatrix}
  \ \pmod 1
  \ ,\quad \delta\in\Sone
  \ .
\end{equation}
On the other hand, there is no anti-symplectic symmetry $\tau$ with
\begin{equation}
  \label{eq:anti-symplectic-symmetries}
  \tau\circ \Kron\circ \tau^{-1}=\Kron^{-1}
  \ ,
\end{equation}
and thus Kronecker maps are not invariant under time reversal.

When discussing the quantisation of the map in the next section, we
will also need a generating function of the classical map. Due to the
independence of the dynamics in 
position and momentum, see~\eqref{eq:kronecker-map}, one
can only give a generating function in a mixed representation,
\begin{equation}
  \label{eq:k-generating-1}
  G_{\alpha\beta}\left(q_{n+1},p_n\right)
  = p_n q_{n+1} + \beta q_{n+1} - \alpha p_n
  \ ,
\end{equation}
with
\begin{equation}
  \label{eq:k-generating-2}
  p_{n+1}
  =\frac{\partial G_{\alpha\beta}\left(q_{n+1},p_n\right)}
  {\partial q_{n+1}}
  \quad \text{and} \quad 
  q_n
  =\frac{\partial G_{\alpha\beta}\left(q_{n+1},p_n\right)}
  {\partial p_n}
  \ .
\end{equation}

 
\rcsInfo $Id: quantum.tex,v 3.2 2001/06/26 10:12:13 guest Exp guest $
\section{Quantum mechanics of Kronecker maps}
\label{sec:quant-mech-kron}
Let us first review the basics of quantum mechanics on the two-torus $\TT$,
see, \eg, \cite{HanBer80,Deg93,DegGraIso95,BieDegGia96,BouBie96}. Phase
space translation operators $\op{T}(q,p)$ yield a representation of
the Heisenberg group,
\begin{equation}
  \label{eq:weyl-heisenberg}
  \op{T}(q,p)\op{T}(q',p')
  = \ue^{\frac{\ui}{2\hbar}(q'p-qp')} \, \op{T}(q+q',p+p')
  \ .  
\end{equation}
Because of the topology of the torus, quantum states $\psi$ have to be
periodic up to a phase in both position and momentum, \ie
\begin{equation}
  \label{eq:periodicity}
  \op{T}(1,0) | \psi \rangle=\ue^{-\zpi \kappa_1} | \psi \rangle
  \quad \text{and}\quad 
  \op{T}(0,1) | \psi \rangle =\ue^{\zpi \kappa_2} | \psi \rangle
  \ .
\end{equation}
For Kronecker maps arbitrary~$\kappa_1$ and~$\kappa_2$ are allowed,
but the spectrum is not changed by choosing different
values. Hence for simplicity we choose $\kappa_1=\kappa_2=0$. The
conditions~\eqref{eq:periodicity} require $\hbar=(2\pi N)^{-1}$ with
$N\in\N$. Moreover one concludes that a quantum state in position (or
in momentum) representation is given by a superposition of
$\delta$-distributions
\begin{equation}
  \label{eq:position-representation}
  \psi(q)
  =\braket{q}{\psi}
  =\frac{1}{\sqN} 
  \sum_{m\in\Z}\sum_{Q\in\ZN} \psi(Q) \ \delta\left(q-\fracN[Q]-m\right)
  \ ,
\end{equation}
with $\ZN\defin\Z/N\Z$. Position states are defined by
\begin{equation}
  \label{eq:position-states}
  \braket{q}{Q}\defin\frac{1}{\sqN}
  \sum_{m\in\Z}\delta\left(q-\fracN[Q]-m\right)
  \ , \quad \text{with} \ Q\in\ZN \ ,
\end{equation}
and momentum states are given by discrete Fourier transformation of
the position states with
\begin{equation}
  \label{eq:scalar-product-position-momentum}
  \braket{Q}{P}=\frac{1}{\sqrt{N}}\ue^{\fracN[\zpi] PQ}
  \ .
\end{equation}  
We remark that now the action of translations in either momentum or
position is given by
\begin{equation}
\begin{aligned}
  \label{eq:translationen}
  \op{T}\left(\tfrac{n}{N},0\right) \ket{Q} &= \ket{Q+n} \, , \quad
  &&\op{T}\left(0,\tfrac{m}{N}\right) \ket{Q} = \ue^{\frac{\zpi}{N} m
    Q} \ket{Q}
  \quad \text{and} \\
  \op{T}\left(\tfrac{n}{N},0\right) \ket{P} &= \ue^{-\frac{\zpi}{N} n
    P} \ket{P} \, , \quad &&\op{T}\left(0,\tfrac{m}{N}\right) \ket{P}
  = \ket{P+m} \, ,
\end{aligned}
\end{equation}
$n,m \in \ZN$. Finally the Hilbert-space $\hilbert\cong\C^N$ on the
two-torus has dimension $N$.  Classical observables 
$f(q,p)\in C^\infty(\TT)$ can be expanded in a Fourier series
\begin{equation}
  \label{eq:classical-observables}
  f(q,p)=\sum_{n,m\in\Z}f_{nm} \, \ue^{\zpi(mq-np)}
  \ ,
\end{equation}
and their Weyl-quantisations are defined as
\begin{equation}
  \label{eq:quantum-observables}
  \Op(f)
  \defin\sum_{n,m\in\Z}f_{nm} \, \op{T}\left(\fracN[n],\fracN[m]\right)
  \ .
\end{equation}

Having reviewed the kinematic aspects of quantised torus maps, the
next step is to quantise the dynamics, \ie to find a suitable quantum
time evolution operator $\op{U}(\Kron)$. This will be
achieved in terms of a Van~Vleck-propagator~\cite{HanBer80} 
with the generating
function~\eqref{eq:k-generating-1}. However, in order to make this
quantisation compatible with the discrete lattice, which the wave
functions are supported on, cf.~\eqref{eq:position-representation},
one first has to find $N$-dependent rational approximations of~$\alpha$ 
and~$\beta$, as has been pointed out by Marklof and
Rudnick for quantum skew maps~\cite{MarRud00}. Following their
approach, we approximate the irrational numbers by fractions with
denominator~$N$, fulfilling
\begin{equation}
  \label{eq:k-approximation}
  \abs{\alpha-\fracN[a]}\leq\frac{1}{2N}
  \quad \text{and}\quad 
  \abs{\beta-\fracN[b]}\leq\frac{1}{2N}
  \ .
\end{equation}
Thus, it is ensured that in the semiclassical limit $N\to\infty$ we
recover the irrational~$\alpha$ and~$\beta$. With the generating
function~\eqref{eq:k-generating-1} and the
approximations~\eqref{eq:k-approximation} we now define the propagator
in a mixed representation by
\begin{equation}
  \label{eq:van-vleck}
  \braopket{Q}{\op{U}(\Kron)}{P}
  \defin\frac{1}{\sqN} 
  \left|\frac{\partial^2 
      G_{\frac{a}{N}\frac{b}{N}}(q,p)}{\partial p\partial q}
  \right|^{\frac{1}{2}}_{\substack{ q=\frac{Q}{N}\\p=\frac{P}{N}}} \,
  \exp\left[{\zpi N \,  
      G_{\frac{a}{N}\frac{b}{N}}\left(\frac{P}{N},\frac{Q}{N}\right)} 
  \right]
  \ ,
\end{equation}
with $Q,P\in\ZN$. Note that by~\eqref{eq:k-generating-1} the
pre-factor is constant. As can easily be verified
using~\eqref{eq:translationen} we also have the identity
\begin{equation}
  \label{eq:k-propagator}
  \op{U}(\Kron)=\op{T}\left(0,\fracN[b]\right)
  \op{T}\left(\vphantom{\fracN[b]}\fracN[a],0\right)
  \ .
\end{equation}
The time evolution operator $\op{U}(\Kron)$ is a quantisation of the
classical map~\eqref{eq:kronecker-map} in the sense that in the
semiclassical limit $N\to\infty$ it fulfils the Egorov property
\begin{equation}
  \label{eq:k-egorov}
  \norm{\op{U}^{-t}(\Kron)\Op(f)\op{U}^t(\Kron)-
    \Op(f\circ \Kron^t)}_{\hilbert}
  =\grosso\left(N^{-1}\right)
\end{equation}
for all observables $f\in C^\infty(\TT)$ and all fixed times~$t\in\N$.
A proof of~\eqref{eq:k-egorov} is given in appendix~\ref{app:egorov}.

The propagator~\eqref{eq:k-propagator} does not preserve all classical
symmetries which were discussed in section~\ref{sec:kronecker-map}.
First of all the Hilbert space structure restricts the continuous
families of translations in~\eqref{eq:k-position-symmetry} 
and~\eqref{eq:k-momentum-symmetry} to discrete translations
\begin{equation}
  \label{eq:k-qm-symmetries-translation-1}
  \op{T}\left(\fracN[c\vphantom{d}],0\right)
  \quad\text{and}\quad
  \op{T}\left(0,\fracN[d]\right)
  \ ,
\end{equation}
with $c,d\in\ZN$. But not all of these translation operators
$\greek{p}\defin\Op(\pi)$ fulfil
\begin{equation}
  \label{eq:unitary-symmetrie}
  \greek{p}\op{U}(\Kron)\greek{p}^{-1}=\op{U}(\Kron)
  \ .
\end{equation}
This is only valid for $\nicefrac{bc}{N}\in\Z$ and
$\nicefrac{ad}{N}\in\Z$ respectively. Consequently, only translations
\begin{equation}
  \label{eq:k-qm-symmetries-translation-2}
  \op{T}\left(\frac{k_b}{D_b},0\right)
  \ ,\quad\text{with}\quad 
  D_b\defin\gcd(b,N)
  \ ,\quad k_b\in\ZN[D_b]
  \ ,
\end{equation}
and
\begin{equation}
  \label{eq:k-qm-symmetries-translation-3}
  \op{T}\left(0,\frac{k_a}{D_a}\right)
  \ ,\quad\text{with}\quad D_a\defin\gcd(a,N)
  \ ,\quad k_a\in\ZN[D_a]
  \ ,
\end{equation}
are symmetries of the quantum propagator~\eqref{eq:k-propagator}.
Beside these quantum symmetries, which correspond directly to the
classical translations~\eqref{eq:k-position-symmetry}
and~\eqref{eq:k-momentum-symmetry}, there are further quantum
symmetries, 
\begin{equation}
  \label{eq:k-qm-symmetries-translation-4}
  \op{T}\left(\fracN[c],\fracN[d]\right)
  \ ,\quad\text{with}\quad 
  \fracN[bc-ad]\in\Z
  \ ,
\end{equation}
which correspond to a combination of classical symmetries.
Obviously, the last class of
symmetries~\eqref{eq:k-qm-symmetries-translation-4} includes the
previous ones. The inversion symmetry~\eqref{eq:k-inversion} is
preserved and the corresponding operator~$\op{I}$ is given by
\begin{equation}
  \label{eq:k-qm-symmetries-inversion}
  \left[\op{I}\psi\right](Q)=\overline{\psi(-Q)}
  \ .
\end{equation}
Summarising we emphasise that, although not all symmetries are
preserved by the quantisation of Kronecker maps, there are quantum
symmetries in position and momentum, which are remnants of the
classical translation invariance. However, it depends on the
rational approximations
for~$\alpha$ and~$\beta$,
and thus also on the semiclassical parameter $N$,
which of these phase
space translations are conserved. Hence the characteristics of the quantum
maps depend on number theoretical properties of~$\alpha$ and~$\beta$.


\rcsInfo $Id: trace.tex,v 3.1 2001/06/26 10:23:35 guest Exp guest $
\section{Trace formula and spectral statistics}
\label{sec:trace-formula}
The purpose of this section is to investigate the spectrum of the
propagator $\op{U}(\Kron)$. The eigenangles~$\vartheta_j$ and the
eigenstates~$\ket{j}$ are defined by
\begin{equation}
  \label{eq:eigenangles}
  \op{U}(\Kron)\ket{j}
  =\ue^{\fracN[\zpi]\vartheta_j}\ket{j}
  \ ,\quad j\in\ZN
  \ .
\end{equation}
Notice that by this definition the eigenangles are already unfolded,
\ie rescaled to mean density one.  One could now explicitly calculate
the eigenangles and the eigenvalues of the Kronecker map.
However, we will not do so but instead discuss spectral properties in
terms of exact and semiclassical trace formulae
similar to the Gutzwiller trace formula~\cite{Gut71}. This procedure
will allow for an immediate generalisation to the case with spin which
will be treated in the following sections. To this end recall that the
spectral density can be expressed by traces of $\op{U}^l(\Kron)$,
\begin{equation}
  \label{eq:spectral-density}
  d(\vartheta)
  \defin \sum_{k\in\Z}\sum_{j\in\ZN}\delta\left(\tfrac{2\pi}{N}
    \left(\vartheta-\vartheta_j\right)-2\pi k\right)
  =\frac{1}{2\pi}\sum_{l\in\Z}\ue^{\fracN[\zpi] l\vartheta}
  \Tr\op{U}^l(\Kron)
  \ .
\end{equation}
Thus, in order to derive a trace formula it is sufficient to calculate
$\Tr\op{U}^l$. One easily verifies by induction that the $l$-step
propagator in mixed representation is given by, 
cf.~\eqref{eq:van-vleck},
\begin{equation}
  \label{eq:l-step-van-vleck}
  \braopket{Q}{\op{U}^l(\Kron)}{P}
  = \frac{1}{\sqN} \exp\left[{\zpi N 
      G^l_{\frac{a}{N}\frac{b}{N}}\left(\frac{P}{N},\frac{Q}{N}\right)} 
  \right] \ ,
\end{equation}
where
\begin{equation}
  \label{eq:l-step-generating}
  G^l_{\alpha\beta}(q,p) 
  = pq + l\beta q - l\alpha p + \frac{l(l-1)}{2} \alpha\beta
\end{equation} 
generates the $l$-step Kronecker map via
\begin{equation}
  \label{eq:l-step-generating-equations}
  p_l = \frac{\partial G^l_{\alpha\beta}(q_l,p_0)}{\partial q_l} 
  \quad  \text{and} \quad
  q_0 = \frac{\partial G^l_{\alpha\beta}(q_l,p_0)}{\partial p_0} \ . 
\end{equation}
In position representation the propagator reads
\begin{equation}
  \label{eq:l-step-propagator-position}
  \braopket{Q}{\op{U}^l(\Kron)}{Q^\prime}
  = \exp\left[ \frac{2\pi\ui}{N} \left( 
      lbQ + \frac{l(l-1)}{2} ab \right) \right]
    \delta_{Q,Q^\prime+la \, \bmod N} \ ,
\end{equation}
and thus the trace is given by
\begin{equation}
  \label{eq:k-trace-1}
  \Tr\op{U}^l(\Kron)
  = N \, \ue^{\frac{2\pi\ui}{N}\frac{l(l-1)}{2}ab} \, 
    \delta_{0,la \, \bmod N} \, \delta_{0,lb \, \bmod N} \ .
\end{equation}
We only obtain a non-zero trace if both~$la$ and~$lb$ are multiples 
of~$N$.  With $\gcd(a,N)=D_a$ and $\gcd(b,N)=D_a$ we define
$M_a=\nicefrac{N}{D_a}$ and $M_b=\nicefrac{N}{D_b}$ respectively.
Hence, in order to get get a non-vanishing trace,
$l$ has to be a multiple of both~$M_a$ and~$M_b$. 

The Kronecker map with translations~$\nicefrac{a}{N}$ 
and~$\nicefrac{b}{N}$ has a two-parameter family of periodic 
orbits~$\gamma$. The periodic points are given by all $(q,p)\in\T^2$ 
and the length of the periodic orbits is determined by
\begin{equation}
  \label{eq:k-length-po}
  l_{\gamma}=\lcm\left(M_a,M_b\right)
  \ .
\end{equation}
Thus the condition for a non vanishing contribution
in~\eqref{eq:k-trace-1} exactly singles out those values of~$l$ which
correspond to multiples of~$l_{\gamma}$.

The exponent of~\eqref{eq:k-trace-1} has to be interpreted as the
action of the periodic orbit in the following sense. In general one
can pass from the generating function $G^l_{\alpha\beta}(q,p)$ in
mixed representation to a generating function
$S^l_{\alpha\beta}(q,q^\prime)$ in position representation by the
Legendre transform
\begin{equation}
  \label{eq:S-action}
  S^l_{\alpha\beta}(q,q^\prime) \defin G^l_{\alpha,\beta}(q,p) - p q^\prime \ .
\end{equation}
Here, this step is only formal, since we cannot solve $q^\prime =
\frac{\partial G^l_{\alpha,\beta}(q,p)}{\partial p}$ for~$p$ in order to
eliminate the initial momentum from~\eqref{eq:S-action}. However, we
can use~\eqref{eq:S-action} to calculate the action
\begin{equation}
\label{eq:l-step-action}
  \fancyS^l \defin S^l_{\frac{a}{N}\frac{b}{N}}
                \left( \frac{Q+la}{N},\frac{Q}{N} \right) 
              - l \frac{b}{N} \frac{Q}{N}
  = \frac{l(l-1)}{2} \frac{a}{N} \frac{b}{N}
\end{equation}
of a periodic orbit, which appears in the trace formula for quantum
maps.  The last term involves the winding number~$lb$ in momentum and
thus accounts for the difference of the action~$\fancyS^l$ on the
torus~$\T^2$ and the action
\begin{math}
  S^l_{\nicefrac{a}{N}\,\nicefrac{b}{N}}
  \left(\frac{Q+la}{N},\frac{Q}{N} \right)
\end{math}
on the covering plane~$\R^2$.  Therefore, the trace of~$\op{U}^l$ is
given by
\begin{equation}
  \label{eq:k-trace-2}
  \Tr\op{U}^l(\Kron) = 
  N \, \ue^{2\pi\ui N \fancyS^l} \, 
    \delta_{0,l \, \bmod M_a} \, \delta_{0,l \, \bmod M_b} 
    \ .
\end{equation}
Finally, we can express the spectral density~\eqref{eq:spectral-density} 
in terms 
of the periodic orbits of the classical map which we used to approximate
the original map for quantisation,
\begin{equation}
  \label{eq:k-trace-formula}
  d(\vartheta)
  =\frac{N}{2\pi}+\frac{N}{2\pi}{\sum_{\substack{k\in\Z \\ k\neq 0}}} 
  \ue^{\frac{\zpi}{N}kl_\gamma\vartheta}
  \ue^{2\pi\ui N \fancyS^{kl_\gamma}}
  \ .
\end{equation}
Thus we have derived an exact trace formula for quantised
Kronecker maps with fixed~$N$.
In the semiclassical limit the approximations~$a$ and~$b$ change and
so does the length~$l_{\gamma}$ of the periodic orbit. For an
irrational translation on the two-torus we have
\begin{equation}
  \label{eq:k-length-po-sc-limit}
  l_{\gamma}\xrightarrow[N\to\infty]{}\infty
  \ ,
\end{equation}
as should be expected for a
uniquely ergodic map without periodic orbits.
\begin{figure}
  \subfigure[Non-universal level spacing distribution of
  the Kronecker map.]  {\includegraphics[width=0.49\textwidth]
    {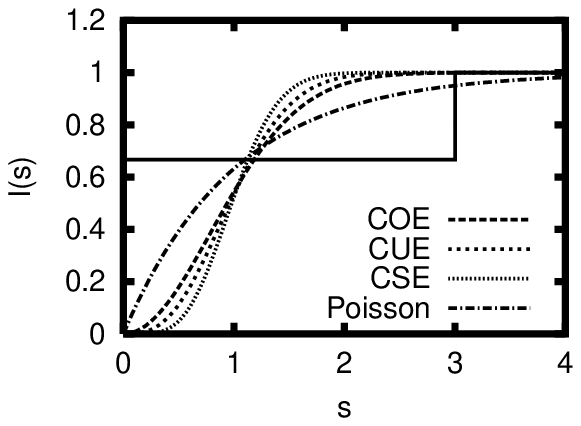}
    \label{fig:kronecker-ergodic-levelspacing}}
  \subfigure[Form factor of the Kronecker map which has peaks at
  $\tau=k\,\nicefrac{l_\gamma}{N}=\nicefrac{k}{3}$ with
  $k\in\Z$.]{\includegraphics[width=0.49\textwidth]
    {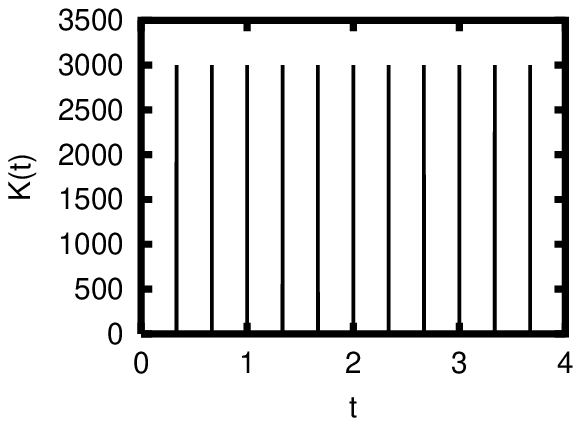}
    \label{fig:kronecker-ergodic-formfactor}}
  \caption{Spectral statistics of a Kronecker map with 
    $\alpha=\nicefrac{(\sqrt{5}-1)}{2}$,
    $\beta=\nicefrac{\sqrt{2}}{2}$ and $N=3000$. The
    approximations~\eqref{eq:k-approximation} lead to $a=1854$ and
    $b=2121$. Thus we get $M_a=500$ and $M_b=1000$.}
  \label{fig:kronecker-ergodic-statistics}
\end{figure}

From the spectral density~\eqref{eq:k-trace-formula} it is clear that
the Kronecker map does not show generic spectral statistics.  Instead
the behaviour for fixed~$N$ depends crucially on the
approximations~\eqref{eq:k-approximation}. This can be observed
explicitly by investigating the form factor
\begin{equation}
  \label{eq:form-factor}
    K(\tau)
    \defin\fracN
    \abs{\Tr\op{U}^{\tau N}\left(\Kron\right)}^2-N\delta_{\tau,0}
    =N\delta_{\tau N,0\;\bmod\;l_\gamma}
    \ ,
\end{equation}
which has peaks if~$\tau$ is a multiple of~$\nicefrac{l_\gamma}{N}$ as
shown in figure~\ref{fig:kronecker-ergodic-formfactor}.  Also the
probability density for the spacings
$s_j\defin\vartheta_{j+1}-\vartheta_{j}$, the level spacing
density~$P(s)$, defined by
\begin{equation}
  \label{eq:level-spacing-distribution}
  \int_a^b P(s) \, \ud s 
  \defin \frac{\#\left\{s_j\in(a,b), j\in\ZN\right\}}{N}
  \ ,
\end{equation}
behaves non-universal. In
figure~\ref{fig:kronecker-ergodic-levelspacing} the
level spacing distribution
\begin{equation}
  \label{eq:integrated-level-spacing-distribution}
  I(s)\defin\int_0^s P(s')\intd s'
  \ 
\end{equation}
is compared to the respective results of the circular orthogonal,
unitary and symplectic ensembles (COE, CUE and CSE respectively) of
RMT and the Poissonian level spacing distribution.  We only remark
that in the limit $N\to\infty$ these statistics have no limit
distribution but depend crucially on number theoretical properties of
the approximations~\eqref{eq:k-approximation}, cf. the similar
situation for the quantised skew map which was discussed in detail
in~\cite{BacHaa99}.  This behaviour is related to the ($N$-dependent)
quantum symmetries~\eqref{eq:k-qm-symmetries-translation-4} which are
remnants of the classical translation
invariance~\eqref{eq:k-position-symmetry}
and~\eqref{eq:k-momentum-symmetry}. The connection becomes evident in
the trace formula~\eqref{eq:k-trace-formula} where all degenerate
periodic orbits collectively contribute to the spectral density.


\rcsInfo $Id: spin.tex,v 3.2 2001/06/26 10:36:23 guest Exp guest $
\section{Kronecker maps with spin -- generic statistics}
\label{sec:maps-with-spin}
Having identified the quantum symmetries (and thus the classical
symmetries) of the Kronecker map as being responsible for its
non-generic spectral statistics, we now aim at breaking these
symmetries without changing the characteristic properties of the
classical system, \ie without introducing any stronger chaotic property 
beyond (unique) ergodicity.
To this end we couple the quantised map to a
spin~$\nicefrac{1}{2}$~\cite{Sch89,KepMarMez01}, \ie we seek a quantum
propagator acting on $\hilbert \otimes \C^2$ instead of~$\hilbert$.
The explicit coupling procedure can be motivated by a Pauli equation
where the translational dynamics is periodically kicked in
time~\cite{KepMarMez01}, yielding
\begin{equation}
\label{eq:Kron_mit_spin}
  \Uspin(\Kron) = 
  \big(\op{U}(\Kron) \otimes \eins_2\big) 
  \Op\left(\ue^{\ui\vecsig\cdot\vecB(q,p)}\right) 
  \ .
\end{equation}
Here~$\eins_2$ is the $2\times2$ unit matrix, the components of
$\vecB(q,p)$ are in $C^\infty(\T^2)$ and~$\vecsig$ denotes the vector
of Pauli matrices with
\begin{equation}
  \sigma_x = \begin{pmatrix} 0 & 1 \\ 1 & 0 \end{pmatrix} \ , \quad
  \sigma_y = \begin{pmatrix} 0 & -\ui \\ \ui & 0 \end{pmatrix} 
  \quad \text{and} \quad
  \sigma_z = \begin{pmatrix} 1 & 0 \\ 0 & -1 \end{pmatrix} \ .
\end{equation}
Thus, $\ue^{\ui\vecsig\cdot\vecB(q,p)}$ takes values in $\SU(2)$ and
is quantised by applying the Weyl
quantisation~\eqref{eq:quantum-observables} to its components. The
physical meaning of $\vecB(p,q)$ is that of a combination of magnetic
fields and spin-orbit coupling terms, and thus $\Uspin(\Kron)$ describes
iterations of the quantum map alternating with spin precession.

Note that the coupling of the spin degrees of freedom to the quantum
map is in non-leading semiclassical order, as can be seen
from~\eqref{eq:Kron_mit_spin}, where
in contrast to~\eqref{eq:van-vleck} 
there is no pre-factor~$N$ in the
exponent of the spin part.
Therefore the corresponding classical dynamics on the torus is not
perturbed but only accompanied by a cocycle taking values in $\SU(2)$
as we will point out in the following.

By multiple applications of the Egorov property~\eqref{eq:k-egorov} in
the semiclassical limit we can write~\cite{KepMarMez01} (by a slight
abuse of notation we define 
$\op{U}(\Kron) \otimes \eins_2 \equiv \op{U}(\Kron)\,$)
\begin{equation}
\label{eq:Uspin_sc}
\begin{split} 
  &\Uspin^t(\Kron)\\
  &\ = \left( \op{U}(\Kron) \Op\left(\ue^{\ui\vecsig\cdot\vecB}
    \right)^{\vphantom{\prime}}
  \right)^t \\
  &\ = \op{U}^t(\Kron) \ \op{U}^{-(t-1)}(\Kron)
  \Op\left(\ue^{\ui\vecsig\cdot\vecB}\right) \op{U}^{(t-1)}(\Kron) \ 
  \op{U}^{-(t-2)}(\Kron) \cdots  \\
  &\ \hspace{27.7ex} \cdots \op{U}^{2}(\Kron) \ \op{U}^{-1}(\Kron)
  \Op\left(\ue^{\ui\vecsig\cdot\vecB}\right) \op{U}(\Kron)
  \Op\left(\ue^{\ui\vecsig\cdot\vecB}\right) \\
  &\ = \left(\op{U}^t(\Kron) \Op\left(\ue^{\ui\vecsig\cdot\vecB}
      \circ \Kron^{t-1} \right) \cdots
    \Op\left(\ue^{\ui\vecsig\cdot\vecB}\circ \Kron \right)
    \Op\left(\ue^{\ui\vecsig\cdot\vecB}\right)\right)
  \left(1+\grosso\left(N^{-1}\right)^{\vphantom{\prime}}\right) \\
  &\ = \op{U}^t(\Kron) \Op\left(d_t\right) 
  \left(1+\grosso\left(N^{-1}\right)^{\vphantom{\prime}}\right) \ ,
\end{split} 
\end{equation} 
where the cocycle~$d_t$ is defined
by~\cite{BolKep98,BolKep99a,KepMarMez01}
\begin{equation}
  \label{eq:cocycle}
  d_t(q,p) \defin 
  \ue^{\ui\vecsig\vecB\left(\Kron^{t-1}(q,p)\right)} \,  
  \ue^{\ui\vecsig\vecB\left(\Kron^{t-2}(q,p)\right)} 
  \cdots \ue^{\ui\vecsig\vecB(q,p)} 
  \ . 
\end{equation}
For later reference we remark that spin dynamics and classical map can
be combined to form a skew product on $\T^2\times\SU(2)$
by~\cite{BolKep99b}
\begin{equation}
\begin{split} 
  \label{eq:skewproduct}
  \begin{array}{crclc}
    Y : 
    & \T^2\times\SU(2) 
    & \longrightarrow 
    & \T^2\times\SU(2) \\[1ex]
    & (q,p,g) 
    & \longmapsto 
    & \left(\Kron(q,p),\, \ue^{\ui\vecsig\vecB(q,p)} g \right) &\ ,
  \end{array}
\end{split} 
\end{equation}
thus implying, $Y^t(q,p,g)= (\Kron^t(q,p), d_t(q,p) g)$. Notice
that~$Y$ leaves the product measure $\mu=\mu_{\T^2} \times \mu_H$
invariant which consists of Lebesgue measure~$\mu_{\T^2}$ on the torus
and the normalised Haar measure~$\mu_H$ on~$\SU(2)$. Observables are
now given by hermitian $2\times2$ matrices
\begin{equation}
  F(q,p) = \begin{pmatrix} f^{11}(q,p) & f^{12}(q,p) \\
                           f^{21}(q,p) & f^{22}(q,p) \end{pmatrix}
\end{equation}
with~$f^{jk} \in \C^\infty(\T^2)$ which are quantised component-wise
by~\eqref{eq:quantum-observables}.  From~\eqref{eq:k-egorov}
and~\eqref{eq:Uspin_sc} now follows the Egorov property for Kronecker
maps with spin,
\begin{equation}
\label{eq:spin-egorov}
  \norm{\Uspin^t \Op(F) \, \Uspin^t - \Op(d_t^{-1}) 
        \Op\left(F\circ\Kron^t\right) \Op(d_t)}_{\hilbert \otimes \C^2}
  =\grosso\left(N^{-1}\right) \ .
\end{equation}
for all fixed times $t \in \Z$.  Clearly, when applying this Egorov
property to a scalar observable, \ie $F(q,p) = f(q,p) \eins_2$, it
reduces to~\eqref{eq:k-egorov}. Thus the classical dynamics on the
torus remains unchanged.

\begin{figure}[!ht]
  \subfigure[Level spacing density in comparison to
  different circular ensembles.]
  {\includegraphics[width=0.49\textwidth]
    {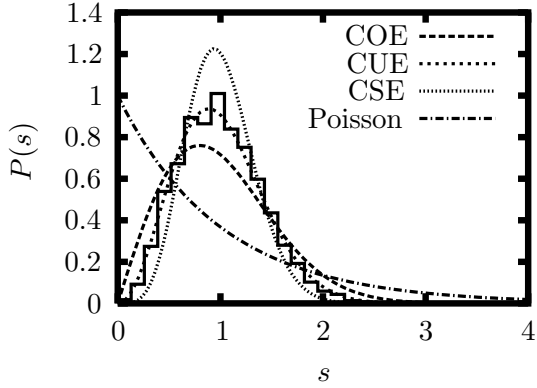}
    \label{fig:kronecker-ergodic-spin-levelspacing-p}}
  \subfigure[In contrast to fig.~\ref{fig:kronecker-ergodic-formfactor} 
  the form factor shows generic fluctuations.]
  {\includegraphics[width=0.49\textwidth]
    {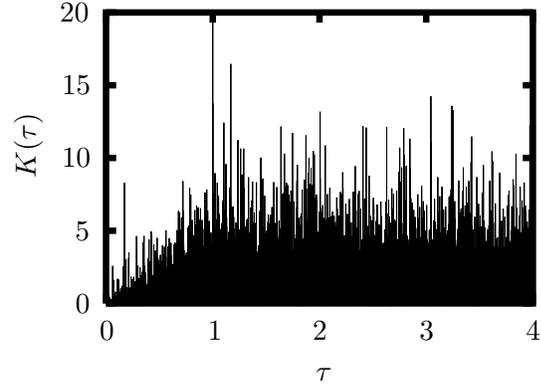}
    \label{fig:kronecker-ergodic-spin-formfactor}}
  \subfigure[The level spacing distribution $I(s)$.]
  {\includegraphics[width=0.49\textwidth]
    {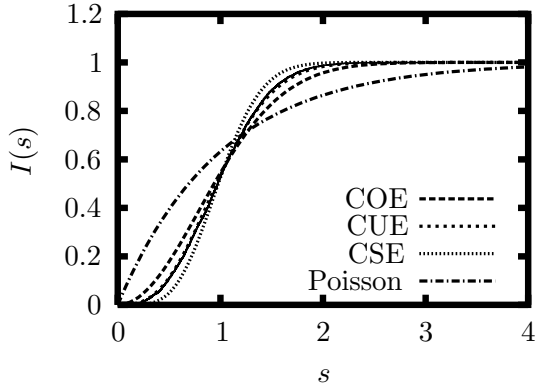}
    \label{fig:kronecker-ergodic-spin-levelspacing}}
  \subfigure[Smoothed form factor where we have averaged over
  $\Delta\tau=0.2$.]  {\includegraphics[width=0.49\textwidth]
    {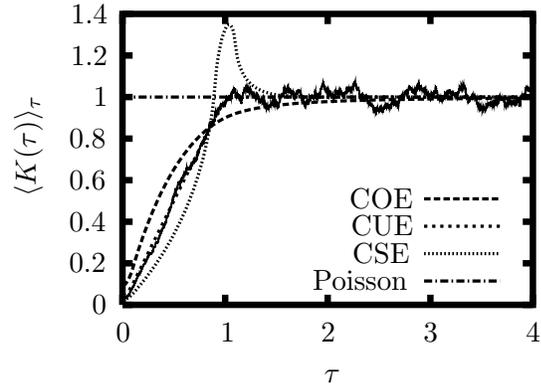}
    \label{fig:kronecker-ergodic-spin-formfactor-mean}}
  \caption{Spectral statistics of the Kronecker map with 
    spin~$\nicefrac{1}{2}$. As in
    fig.~\ref{fig:kronecker-ergodic-statistics} we have chosen
    $\alpha=\nicefrac{(\sqrt{5}-1)}{2}$,
    $\beta=\nicefrac{\sqrt{2}}{2}$ and $N=3000$. Thus the Hilbert
    space has dimension $2N=6000$. We observe good agreement with the 
    CUE.}
  \label{fig:kronecker-spin-ergodic-statistics}
\end{figure}

We now present some numerical results on the spectral statistics of
the Kronecker map with spin.  For convenience instead of
$\Op\left(\ue^{\ui\vecsig\vecB(q,p)}\right)$ we choose
\begin{math}
  \Op\left(\ue^{\ui\vecsig\vecB_1(q)}\right) \, 
  \Op\left(\ue^{\ui\vecsig\vecB_2(p)}\right)
\end{math}
with
\begin{equation}
\label{eq:B-Felder}
  \vecB_1(q) = 
  \begin{pmatrix} 
    \sin(2\pi q) \\
    \sin(4\pi q) \\
    \sin(6\pi q)
  \end{pmatrix}
  \quad \text{and} \quad 
  \vecB_2(p) = 
  \begin{pmatrix} 
    \sin(2\pi p) \\
    \sin(4\pi p) \\
    \sin(6\pi p)
  \end{pmatrix} 
  \ .
\end{equation}
Due to the group property of $\SU(2)$ we can always find a unique
$\vecB(q,p)$ with $\ue^{\ui\vecsig\vecB(q,p)} =
\ue^{\ui\vecsig\vecB_1(q)} \, \ue^{\ui\vecsig\vecB_2(p)}$ and by the
Moyal product of the Weyl quantisation, see, \eg, \cite{Fol89}, we have
\begin{equation}
  \Op\left(\ue^{\ui\vecsig\vecB(q,p)}\right)
  = \Op\left(\ue^{\ui\vecsig\vecB_1(q)}\right) 
  \Op\left(\ue^{\ui\vecsig\vecB_2(p)}\right)
    \, \left(1+\grosso\left(N^{-1}\right)^{\vphantom{\prime}}\right) 
    \ .
\end{equation}
Therefore, the above semiclassical discussion applies to the numerical
situation. Notice that the choice of the fields~\eqref{eq:B-Felder}
breaks all symmetries of the quantum map $\op{U}(\Kron)$. However,
since the classical dynamics on the torus is not changed by the
spin-coupling, being of higher order in~$N$, the quantum map $\Uspin(\Kron)$
still corresponds to a non-mixing classical system. If the 
fields~$\vecB_1$ and~$\vecB_2$ generate a sufficiently random dynamics on
$\SU(2)$ we may expect the skew product~\eqref{eq:skewproduct} to be
ergodic, but it can never be mixing since this would also imply the
dynamics on the torus, given by~$\Kron$, to be mixing.

In figure~\ref{fig:kronecker-spin-ergodic-statistics} we show the
level spacing density and the 
level spacing distribution for a
Kronecker map with spin, observing good agreement with the CUE.
This is
consistent with the fact that $\Uspin(\Kron)$ has 
neither unitary nor anti-unitary
symmetries.
The form factor~\ref{fig:kronecker-ergodic-spin-formfactor} shows
typical fluctuations and when averaging (over an interval of length
$\Delta\tau=0.2$ with unit weight) one obtains again nice agreement
with the CUE.

Thus we have provided an example in favour of the BGS conjecture,
where only the weakest possible chaotic property, \ie ergodicity, is
fulfilled. The only difference to quantised Kronecker maps without
spin, which show non-generic spectral statistics, is the absence of
quantum mechanical degeneracies, which
were discussed in the previous sections.  
In the following, we will explain that this is related to the 
lifting of degeneracies of periodic orbits in the semiclassical 
trace formula.
We remark that for different values of 
$\alpha,\beta \in \Sone \smallsetminus \Q$ we 
make similar observations.

 
\rcsInfo $Id: trace-spin.tex,v 3.2 2001/06/26 12:00:54 guest Exp guest $
\section{Trace formula with spin}
\label{sec:trace-formula-with-spin}
The purpose of this section is to derive a trace formula for Kronecker
maps with spin. Semiclassical trace formulae for flows with spin have
been derived in~\cite{BolKep98,BolKep99a} and trace formulae for
(perturbed) cat maps with spin have been investigated
in~\cite{KepMarMez01}.

First recall~\eqref{eq:Uspin_sc} that the propagator of the Kronecker
map with spin fulfils
\begin{equation}
  \label{eq:propagator-with-cocycle}
  \Uspin^l(\Kron)=\op{U}^l(\Kron) \Op\left(d_l\right) \ 
  \left(1+\grosso\left(N^{-1}\right)^{\vphantom{\prime}}\right)
  \ ,
\end{equation}
where~$d_l$ denotes the cocycle~\eqref{eq:cocycle}.  Calculating the
trace of~\eqref{eq:propagator-with-cocycle} thus reduces to deriving a
(semiclassical) trace formula for $\op{U}^l(\Kron)$ with matrix
elements of the quantum observable $\Op(f)$, $f=\tr d_l$, 
where~$\tr(\cdot)$ denotes the trace on~$\C^2$, \ie we have
\begin{equation}
  \label{eq:k-trace-observables-1}
  \Tr\left(\op{U}^l(\Kron)\Op(f)\right)
  =\sum_{Q,Q'\in\ZN}
  \braopket{Q}{\op{U}^l(\Kron)}{Q'}
  \braopket{Q'}{\Op(f)^{\vphantom{l}}}{Q}
  \ .
\end{equation}
With the Weyl quantisation~\eqref{eq:quantum-observables} and the
propagator~\eqref{eq:l-step-propagator-position} we obtain
\begin{equation}
  \label{eq:k-trace-observables-3}
  \Tr\left(\op{U}^l(\Kron)\Op(f)\right)
  =N\sum_{n,m\in\Z}f_{nm}\ue^{\fracN[\ui \pi] \left(mn-l(l-1)ab\right)}
  \delta_{n,la\bmod N}\delta_{m,lb\bmod N}
  \ .
\end{equation}
and thus get a non vanishing trace for
\begin{equation}
  \label{eq:k-trace-observables-condition}
  n\equiv la \bmod N \quad\text{and}\quad m\equiv lb \bmod N
  \ .
\end{equation}
Let us discuss the first condition for fixed $l\neq 0$ in the
semiclassical limit $N\to\infty$. We distinguish the following two cases:
\begin{enumerate}[(i)]
\item \label{item:trivial-condition}$la \equiv 0 \bmod N$:\\
  Then~\eqref{eq:k-trace-observables-condition} is clearly fulfilled
  for $n=0$ which leads to the same condition as without spin,
  see~\eqref{eq:k-trace-2}. The next values of $n$ for
  which~\eqref{eq:k-trace-observables-condition} holds are given by
  $n=\pm N$. In consequence the smallest non-trivial value of n is
  given by $\min\limits_{n\neq0}\abs{n}=N$.
\item \label{item:non-trivial-condition}$la \not\equiv 0 \bmod N$:\\
  The smallest value of $n$ for
  which~\eqref{eq:k-trace-observables-condition} is fulfilled is given
  by $\min\abs{n}=\min(la,N-la)$. With the
  approximation~\eqref{eq:k-approximation} we get
  $\min\abs{n}=\min(lN\alpha+\grosso(1),N(1-l\alpha)+\grosso(1))$.
  Since $l$ is fixed this leads to $\min\abs{n}=\grosso(N)$.
\end{enumerate}
From~\eqref{item:trivial-condition}
and~\eqref{item:non-trivial-condition} it is clear that in the
semiclassical limit the first non-trivial~$n$ which yields a
contribution to~\eqref{eq:k-trace-observables-condition} is given by
\begin{equation}
  \label{eq:non-trivial-n}
  \min_{n\neq 0}\abs{n}=\grosso(N)
  \ .
\end{equation}
The second condition in~\eqref{eq:k-trace-observables-condition} is
discussed analogously. Hence in the
semiclassical limit~\eqref{eq:k-trace-observables-3}
 reduces to the term with $n=m=0$. All other terms
are of higher semiclassical order since $f\in C^\infty(\TT)$ implies
that the coefficients~$f_{nm}$ of the Fourier series decrease
exponentially. We therefore have
\begin{equation}
  \label{eq:k-trace-observables-4}
  \Tr\left(\op{U}^l(\Kron)\Op(f)\right)
  \sim N f_{00} \, 
  \ue^{\fracN[\ui\pi]l(1-l)ab} \, \delta_{0,l \bmod M_a} \, 
  \delta_{0,l \bmod M_b} 
  \ ,
\end{equation}
which can be written in terms of the periodic orbits of the
approximated Kronecker map with length $l_\gamma=\lcm(M_a,M_b)$ and
action~$\fancyS^l$ from equation~\eqref{eq:l-step-action} yielding
\begin{equation}
  \label{eq:k-trace-spin}
  \Tr\left(\Uspin^l(\Kron)\right)
  \sim N \sum_{k\in\Z}\delta_{l,kl_{\gamma}} \, 
  \ue^{\zpi N \fancyS^l}
  \iint_{\TT} \tr d_l(q,p) \dsymplectic
  \ ,
\end{equation}
where we have also substituted the observable~$f(q,p)$ by $\tr
d_l(q,p)$.
As in the case without spin~\eqref{eq:k-trace-2} we still have a
two-parameter family of periodic orbits with
action~$\fancyS^{kl_\gamma}$.  However, the contribution of each
periodic point $(q,p)\in\T^2$ is weighted by the spin trace $\tr
d_{kl\gamma}(q,p)$, thus breaking the degeneracies semiclassically
without changing the periodic orbit structure of the classical torus
map.

\begin{figure}[!ht]
  \subfigure[For lengths $L\ll \THBS$ we get for both maps Gaussian
  peaks only for $L=kl_{\gamma}$, $k\in\Z$. The amplitude of the peaks
  differs for the map with and without spin~$\nicefrac{1}{2}$ due to
  the contribution of the cocycle in the trace
  formula~\eqref{eq:k-trace-spin}.]
  {\includegraphics[width=0.99\textwidth]
    {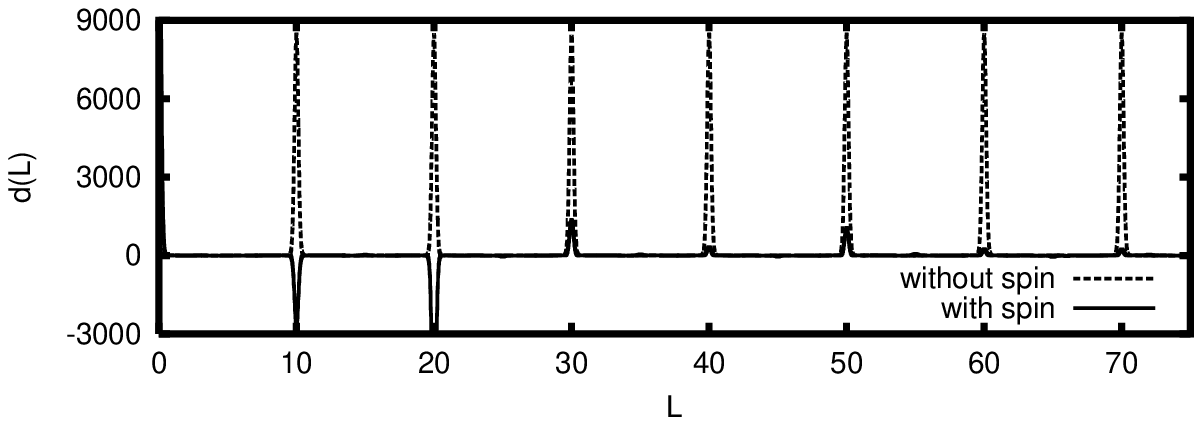}\label{fig:cosheat-small}}
    \subfigure [If $L$ becomes larger we get 
    further peaks due to quantum fluctuations for the map with spin,
    but for 
    $L<\THBS$ the peaks at $L=kl_{\gamma}$ are still dominating.]
  {\includegraphics[width=0.50\textwidth]
    {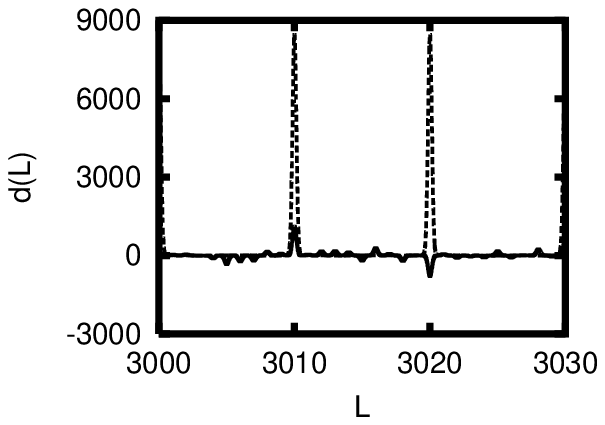}\label{fig:cosheat-medium}}\hspace{-3mm}
  \subfigure[Finally, for lengths of the order $\THBS$ one gets  pure 
  quantum fluctuations for the Kronecker map with spin.]  
  {\includegraphics[width=0.50\textwidth]
    {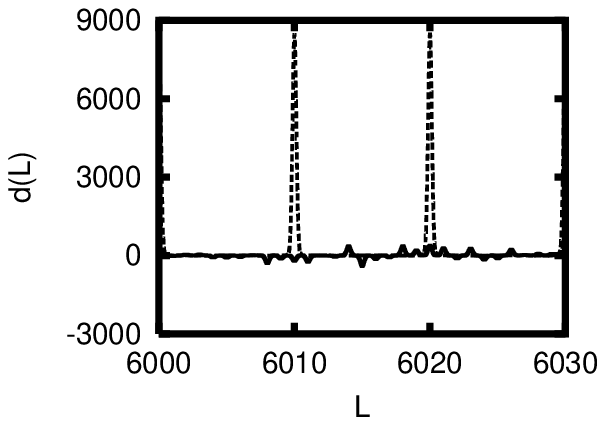}\label{fig:cosheat-large}}
  \caption{Cosine-modulated heat-kernel for a Kronecker map with and 
    without spin. In both cases we have chosen $N=3000$. Thus
    the Heisenberg time is given by $\THB=3000$ and $\THBS=6000$
    respectively. For both maps the approximations of $\alpha$ and
    $\beta$ yield $M_a=5$ and $M_b=2$. Therefore the lengths of the
    periodic orbits are given by multiples of $l_\gamma=10$.}
  \label{fig:cosheat}
\end{figure}

From the Egorov property~\eqref{eq:spin-egorov} it was already clear
that the spin does not change the classical dynamics on the torus.
This can now be nicely illustrated using the trace formula from above.
To this end we evaluate the spectral density on a test-function
$g_\varepsilon(L,\vartheta)$,
\begin{equation}
  \label{eq:smeared-spectral-density}
  d_\varepsilon(L)
  \defin\int_{-\infty}^\infty g_\varepsilon(L,\vartheta) 
  d(\vartheta)\intd\vartheta
  \ .
\end{equation}
Choosing
\begin{equation}
  \label{eq:cos-mod-heat-testfkt}
  g_\varepsilon(L,\vartheta)
  \defin\cos(\vartheta L) \, \ue^{-\vartheta^2 \varepsilon}
  \ ,
\end{equation}
we obtain the so-called trace of the cosine-modulated
heat-kernel~\cite{AurSieSte88,AurSte92}, and with the
identity~\eqref{eq:spectral-density} $d_\varepsilon(L)$ becomes
\begin{equation}
  \label{eq:cos-mod-heat}
  d_\varepsilon(L)
  =\sqrt{\frac{\pi}{\varepsilon}}
  \left(\frac{N}{2\pi}\ue^{-\frac{L^2}{4\varepsilon}}
    +\frac{1}{4\pi}\sum_{l=1}^\infty 
    \ue^{-\frac{(l-L)^2}{4\varepsilon}}\Tr\op{U}^l\right)
  \ .
\end{equation}
Therefore, by~\eqref{eq:k-trace-spin}, in the semiclassical limit
$d_\varepsilon(L)$ has Gaussian peaks at multiples of the periods of
periodic orbits. These can in turn be determined from the quantum
eigenvalues.  For a strongly chaotic map this is rather uninteresting
since for each possible length $L\in\N$ an exponentially increasing
number of orbits contributes to the amplitude.  In our situation,
however, there are only contributions for lengths $L$ which are
multiples of $l_\gamma$. Thus, each peak corresponds to a particular
number $k$ of repetitions of the family $\gamma$ of periodic orbits
and we can observe how spin changes their effective multiplicity, \ie 
the weight factor which these orbits contribute to the trace formula.

In figure~\ref{fig:cosheat} we compare the cosine-modulated
heat-kernel~\eqref{eq:cos-mod-heat} for the Kronecker map with and
without spin~$\nicefrac{1}{2}$. For both maps we have chosen the same
parameters~$\alpha$ and~$\beta$ as well as the same value of~$N$. Thus
we get identical classical dynamics on the torus for the approximating
maps which were used for quantisation. The dimension of the Hilbert
space of the map with spin is twice the dimension of the Hilbert space
of the map without spin.
Thus the Heisenberg time~$\THBS$ for the map with spin is twice the
Heisenberg time $\THB=N$ for the map without spin and for the case
with spin in~\eqref{eq:cos-mod-heat} we simply have to replace~$N$
by~$2N$.

For small~$L$ we get pronounced peaks at multiples of~$l_\gamma$ for
both the map with and without spin. In agreement with the exact trace
formula~\eqref{eq:k-trace-formula} for the pure Kronecker map without 
spin this remains true for all~$L$.  For $\Uspin(\Kron)$,
however, when increasing~$L$ we first observe that the height of the
peaks decreases due to the spin weight $\iint_{\T^2} \tr d_l \, \ud q
\ud p$. Then, for $L \lesssim \THBS$, additional contributions appear
which correspond to higher order corrections to the trace
formula~\eqref{eq:k-trace-spin}.  Finally, for $L \gg \THBS$, these
corrections dominate the behaviour. Summarising we have confirmed that
in the semiclassical regime, \ie for $L \ll \THBS$, the coupling of
the map to a spin only changes the amplitudes of the periodic orbit
contributions but does not affect the classical dynamics on the
torus.

A similar discussion now applies to the spectral form factor, which for
$\Uspin(\Kron)$ is defined by
\begin{equation}
  \label{eq:form-factor-w-spin}
  K(\tau) \defin \frac{1}{\THBS} 
             \left| \Tr \left[ \Uspin(\Kron) \right]^{\tau\THBS} \right|^2
             - 2N \delta_{\tau,0} \ .
\end{equation}
The smallest value of~$\tau$ for which $K(\tau)$ does not vanish is
given by
\begin{equation}
  \label{eq:min-tau}
  \tau_{\mathrm{min}} = \frac{l_\gamma}{\THBS} 
  = \frac{\lcm(M_a,M_b)}{\THBS} \ .
\end{equation}
If~$\alpha$ is irrational then $M_a\to\infty$ in the semiclassical
limit $N\to\infty$~\cite[lemma 4.2]{MarRud00}, and similarly for $M_b$.
Thus if~$\tau$ tends to zero such that $\tau < \tau_{\mathrm{min}}$
one concludes that
\begin{equation}
  \label{eq:k-form-factor-small-tau}
  \lim_{\substack{N\to\infty\\\tau\to0}}K(\tau)=0
  \ .
\end{equation}
This is, however, the same result as we would have obtained in the
case without spin, cf. section~\ref{sec:trace-formula}, which is
consistent with both the non-generic behaviour in
figure~\ref{fig:kronecker-ergodic-formfactor} and the generic
behaviour in figure~\ref{fig:kronecker-ergodic-spin-formfactor-mean}.
It is interesting to remark that with present methods, even in the
regime which is usually dominated by the diagonal
approximation~\cite{HanOzo84,Ber85}, it is not possible to distinguish
semiclassically between generic and non-generic behaviour. This
unusual observation can be traced back to the fact that the Kronecker
map has no periodic orbits and thus in the semiclassical limit,
paradoxically, the periodic orbit contributions to the semiclassical
trace formula do not determine the behaviour of $K(\tau)$ in the
vicinity of $\tau=0$. Conversely, for small but non-zero~$\tau$,
in the semiclassical limit the form factor is dominated by corrections to
the trace formula~\eqref{eq:k-trace-spin} which manifest themselves in
figure~\ref{fig:cosheat} as additional peaks.


\rcsInfo $Id: skew.tex,v 3.1 2001/06/26 11:37:38 guest Exp guest $
\section{Ergodic skew maps}
\label{sec:skew-map}
Another interesting class of parabolic maps on the two-torus~$\TT$ are
the so-called skew maps. We will consider skew maps of the form
\begin{equation}
  \label{eq:skew-map}
  \begin{array}{ccccc}
    \Skew:&\TT&\longrightarrow&\TT& \\[1ex]
    &
    \begin{pmatrix}
      p \\
      q \\
    \end{pmatrix}
    &\longmapsto&
    \begin{pmatrix}
      p+\beta \\
      q+2p \\
    \end{pmatrix}
    &\pmod 1 \ .
  \end{array}
\end{equation} 
For $\beta\in\Sone\smallsetminus\Q$ one has again a uniquely ergodic but
non-mixing map, for details see~\cite{Fur61,AbrRoh65,CouHam97}.
If~$\beta$ is rational the map is in general almost integrable. This
situation will be discussed in the following section, and is our main
motivation for also analysing skew maps in addition to the simplest
case of parabolic maps, namely translations of the torus.  As in the
case of Kronecker maps symmetries play an important r\^ole.  Skew maps
also have several symplectic symmetries. First of all there is a
translation in momentum
\begin{equation}
  \label{eq:s-momentum-symmetry}
  T_p:
  \begin{pmatrix}
    p \\
    q \\
  \end{pmatrix}
  \longmapsto
  \begin{pmatrix}
    p+\frac{1}{2} \\
    q \\
  \end{pmatrix}
  \ \pmod 1 
  \ ,
\end{equation}
and a family of translations in position
\begin{equation}
  \label{eq:s-position-symmetry}
  T_q:
  \begin{pmatrix}
    p \\
    q \\
  \end{pmatrix}
  \longmapsto
  \begin{pmatrix}
    p \\
    q+\gamma \\
  \end{pmatrix}
  \ \pmod 1 
  \ ,\quad \gamma\in\Sone
  \ .
\end{equation}
Moreover, skew maps are invariant under two anti-symplectic
transformations,
\begin{equation}
  \label{eq:s-anti-symplectic-symmetry}
  \tau_{\phantom{p}}:
  \begin{pmatrix}
    p \\
    q \\
  \end{pmatrix}
  \longmapsto
  \begin{pmatrix}
    -p+\delta \\
    q \\
  \end{pmatrix}
  \ \pmod 1 
  \ ,\quad 2\delta\equiv2\beta \bmod 1 
  \ ,
\end{equation}
Thus, in contrast to Kronecker maps, we now have a (non-conventional)
time reversal symmetry.

\begin{figure}[!t]
  \subfigure[The level spacing distribution for a skew map is shown.
  The distribution is obviously non generic.
  The level spacing density is given by
  $P(s)=\nicefrac{1}{3}(\,\delta(s)+\delta(s-1)+\delta(s-2)\,)$, 
  see~\cite{BacHaa99}.] {\includegraphics[width=0.49\textwidth]
    {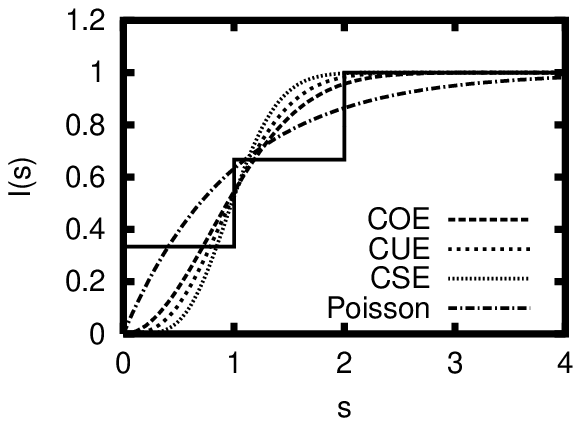}\label{fig:skew-ergodic-levelspacing}}
  \subfigure[The form factor has peaks for
  $\tau=\nicefrac{k}{D}=\nicefrac{k}{3}$, $k\in\N$. For multiples of
  the Heisenberg time $\THB=N$ one gets peaks of the height $N$. The
  average for $\tau>1$ gives a mean degeneracy for the eigenangles of
  $\nicefrac{5}{3}$.]  {\includegraphics[width=0.49\textwidth]
    {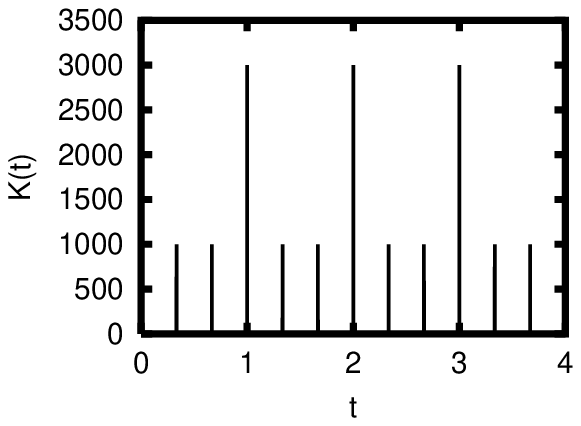}\label{fig:skew-ergodic-formfactor}}
  \caption{Spectral statistics for a skew map with 
    $\beta=\nicefrac{(\sqrt{5}-1)}{2}$ and $N=3000$, implying
    $D=\gcd(b,N)=3$ and $M=\nicefrac{N}{D}=1000$.}
  \label{fig:skew-ergodic-statistics}
\end{figure}

The first quantisation of skew maps, which fulfils an Egorov property
for all $N\in\N$, cf.~\eqref{eq:k-egorov}, was given by Marklof and
Rudnik~\cite{MarRud00}. This quantisation
is similar to the quantisation of the Kronecker map which we presented
in section~\ref{sec:quant-mech-kron}. First of all one has to
approximate the translation in momentum by a rational number
$\nicefrac{b}{N}$ with
\begin{equation}
  \label{eq:s-approximation}
  \abs{\beta-\fracN[b]}\leq\frac{1}{2N} \ , \quad b \in \Z
  \ .
\end{equation}
The propagator $\op{U}(\Skew)$ can then be defined similarly
to~\eqref{eq:van-vleck} by using a generating function
$G(q_{n+1},p_{n})$. In mixed representation one obtains
\begin{equation}
  \label{eq:s-propagator-mixed}
  \braopket{Q}{\op{U}(\Skew)}{P}
  =\frac{1}{\sqrt{N}}
  \exp\left(\zpi N \, G_{\fracN[b]}
    \left(\fracN[Q],\fracN[P]\right) \right)
  =\frac{1}{\sqrt{N}} \, 
  \ue^{\fracN[\zpi]\left(-P^2+Q(P+b)\right)}
  \ ,
\end{equation}
and in momentum representation $\op{U}(\Skew)$ is simply given by
\begin{equation}
  \label{eq:s-propagator-momentum}
  \braopket{P}{\op{U}(\Skew)}{P^\prime}
  = \ue^{-\fracN[\zpi]P^{\prime 2}}\ \delta_{P,P'+b} \ .
\end{equation}
For this propagator the Egorov property was proven in~\cite{MarRud00}.
Again the quantisation does not preserve all classical symmetries.
For our purposes it is not necessary to give a complete list of all
quantum symmetries, instead we only remark that there are always
unitary quantum symmetries given by translations in position
\begin{equation}
  \label{eq:s-qm-symmetrie-position}
  \op{T}\left(\frac{k}{D},0\right)
  \ ,\quad\text{with}\quad D\defin\gcd(b,N)
    \ ,\quad k\in\ZN[D]
  \ ,
\end{equation}
and an anti-unitary symmetry~$\greek{t}\defin\Op(\tau)$ 
involving a translation in
momentum
\begin{equation}
  \label{eq:s-qm-symmetrie-momentum}
  \left[\greek{t}\psi\right](P)=\overline{\psi\left(-P+\fracN[b]\right)}
  \ .
\end{equation}

\begin{figure}[!t]
  \subfigure[The level spacing distribution agrees with the
  CUE.]
  {\includegraphics[width=0.49\textwidth]
    {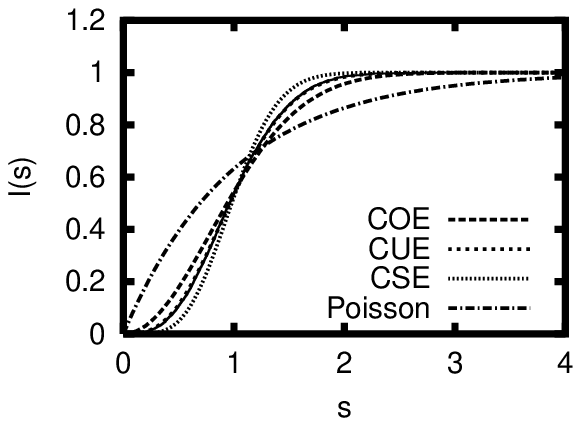}
    \label{fig:skew-spin-ergodic-levelspacing}}
  \subfigure[Smoothed form factor, where we averaged over
  $\Delta\tau=0.2$.]  {\includegraphics[width=0.49\textwidth]
    {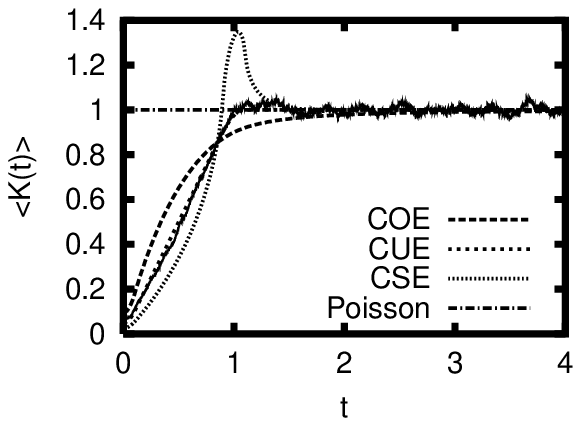}
    \label{fig:skew-spin-ergodic-formfactor}}
  \caption{Spectral statistics for a skew map with spin~$\nicefrac{1}{2}$. 
    The parameters are identical to those 
    chosen in figure~\ref{fig:skew-ergodic-statistics}, \ie
    $N=3000$, $\beta=\nicefrac{(\sqrt{5}-1)}{2}$, $D=3$
    and $M=1000$.}
  \label{fig:skew-spin-ergodic-statistics}
\end{figure}

As for Kronecker maps one can calculate the eigenvalues and
eigenvectors of the propagator~\eqref{eq:s-propagator-mixed}
explicitly~\cite{MarRud00}. Analysing the spectrum of skew maps one
observes non-generic spectral statistics determined by the value of
$D$ defined in~\eqref{eq:s-qm-symmetrie-position}. Furthermore, there is
even no limit distribution~\cite{BacHaa99}. The non-generic behaviour
of the spectral statistics of a quantum skew map is illustrated in
figure~\ref{fig:skew-ergodic-statistics}. For a semiclassical analysis
we again use a trace formula. Details of the derivation can be found
in appendix~\ref{app:skew-trace}. With~\eqref{eq:app-trace-skew-spin-7},
see also~\eqref{eq:app-trace-2}, we obtain for the spectral density
\begin{equation}
  \label{eq:skew-trace-formula}
  d(\vartheta)
  =\frac{N}{2\pi}
  +\frac{\sqrt{N}}{2\pi}\sum_{k\in\Z\smallsetminus\{0\}} \ \sum_{\nu\in\Z_{2k}}
  \frac{M}{\sqrt{2\ui kM}}\,
  \ue^{\frac{\zpi}{N}kM\vartheta} \, 
  \ue^{\zpi N \fancyS^{kM}_\nu}
    \ ,
\end{equation}
with $M\defin\nicefrac{N}{D}$.
The periods of periodic orbits of the approximating map
are given by $kM$,
and $\nu$ labels families of periodic points sharing the same action.
These families form invariant manifolds
$\E_{\nicefrac{\nu}{2kM}}(\nicefrac{b}{N})$,
see~\eqref{eq:disjoint-circles} below, which in particular are
invariant under translations in position as is the classical map.
Again we have high degeneracies in both the quantum spectrum and in
the the trace formula which are the origin of the non-generic spectral
statistics of the quantum map (see
figure~\ref{fig:skew-ergodic-statistics}).

As in the case of Kronecker maps we couple the quantised skew map
to a spin~$\nicefrac{1}{2}$ in order to break these symmetries. As we
pointed out above the skew map also has symmetries in both position
and momentum. Therefore, we use the same magnetic field~$\vecB$ as for
the Kronecker map~\eqref{eq:B-Felder}, depending on both~$q$ and~$p$.
The same calculation as in section~\ref{sec:maps-with-spin} 
then yields the Egorov property for skew maps with spin.

In appendix~\ref{app:skew-trace} we also derive a semiclassical
trace-formula for the skew map with spin.
With~\eqref{eq:app-trace-skew-spin-7} the spectral density reads,
\begin{equation}
  \label{eq:skew-spin-trace-formula}
  d(\vartheta)
  \sim\frac{N}{\pi}
  +\frac{\sqrt{N}}{2\pi}\sum_{k\in\Z\smallsetminus\{0\}} \
  \sum_{\nu\in\Z_{2k}}
  \frac{ \ue^{\frac{\zpi}{N}kM\vartheta} \, 
         \ue^{2\pi\ui N\fancyS^{kM}_\nu}}
       {\sqrt{2\ui kM}} \sum_{j\in\Z_M}
  \int_0^1 
    \tr d_{kM} \!\left(q,\tilde{p}^{(\nu)}+\frac{j}{M}\right) \intd q 
  \ ,
\end{equation}
where now in contrast to~\eqref{eq:skew-trace-formula} the different 
periodic orbits of one family, labelled by~$\nu$, enter with different 
spin weights~$\tr d$. Hence the
magnetic field again breaks the degeneracies in the periodic orbit
sum~\eqref{eq:skew-spin-trace-formula}, but does not change the
classical dynamics on the torus in the sense that the set of periodic
orbits remains the same and only their amplitudes are altered.

In figure~\ref{fig:skew-spin-ergodic-levelspacing} we show the
level spacing distribution and the form factor for an
ergodic skew map with spin~$\nicefrac{1}{2}$.  As in the case of the
Kronecker map we observe good agreement with the CUE. Hence, we have a
further example of a system which is only ergodic, but obeys the
BGS conjecture.

By using the same arguments as in
section~\ref{sec:trace-formula-with-spin} and the semiclassical trace
formula~\eqref{eq:skew-spin-trace-formula} we conclude that for the
form factor of ergodic skew maps
\begin{equation}
  \label{eq:s-form-factor-small-tau}
  \lim_{\substack{N\to\infty\\\tau\to0}}K(\tau)=0
\end{equation}
holds. But again, as in the case of Kronecker maps, the behaviour
of $K(\tau)$ for small but non-zero~$\tau$ cannot simply be determined 
by a semiclassical analysis.

 
\rcsInfo $Id: pseudoint.tex,v 3.2 2001/06/26 12:01:13 guest Exp guest $
\section{Almost integrable skew maps}
\label{sec:skew-map-pseudo}
We now turn our attention to the interesting case of rational skew
maps
\begin{equation}
  \label{eq:skew-map-pseudo}
  \begin{array}{ccccc}
    \Skew:&\TT&\longrightarrow&\TT& \\[1ex]
    &
    \begin{pmatrix}
      p \\
      q \\
    \end{pmatrix}
    &\longmapsto&
    \begin{pmatrix}
      p+\beta\\
      q+2p \\
    \end{pmatrix}
    &\pmod 1 \ ,
  \end{array}
\end{equation} 
with $\beta \in \Sone\cap\Q$, 
\ie we can write $\beta = B/M$ were $M\in\N$ and $B\in\{0,1,\dots,M-1\}$ 
are relatively prime. One immediately finds that invariant sets
of~$S_\beta$ are given by~$M$ disjoint circles
\begin{equation}
  \label{eq:disjoint-circles}
  \E_{p_0}(\beta) \defin \left\{ 
    (q,p) \in \T^2 \ \Big| \ p = p_0 + \frac{j}{M} \, , \ j \in \Z_M 
  \right\} \, .
\end{equation}
If $\beta=0$ then $\E_{p_0}(\beta) \cong \Sone$ and thus the
map is integrable. 
In all other cases one has
$M>1$ and the map restricted to $\E_{p_0}(\beta)$
can be represented as an interval-exchange transformation 
(for interval-exchange transformations see, \eg,
\cite{katokhasselblatt,hasselblattkatok,cornfeldfominsinai} and
references therein).
Therefore, we will call this situation almost integrable 
or pseudointegrable.
We only remark that for $M\to\infty$ we approach
the situation of the last section, \ie when increasing~$M$, starting
from $M=1$, we observe a transition from integrable to ergodic.

For the quantisation of~$S_\beta$ we can use the same procedure as in
section~\ref{sec:skew-map}, with the important difference that we will
not approximate~$\beta$ but instead restrict~$N$ to multiples of~$M$,
\ie $N=DM$, $D\in\N$. As a consequence, the Egorov property,
cf.~\eqref{eq:k-egorov}, is not merely an asymptotic relation but an
identity. Following the notation of section~\ref{sec:skew-map} we will
also write $\beta=\nicefrac{b}{N}$ with $b=BD$.

\begin{figure}
  \subfigure[Level spacing distribution for the map
  where one can see strong degeneracies in the spectrum.]
  {\includegraphics[width=0.49\textwidth]
    {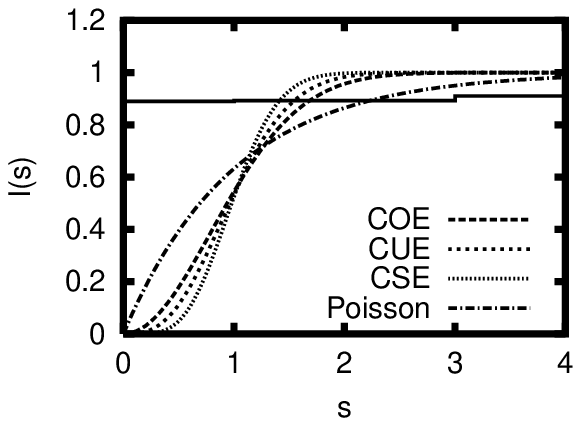}\label{fig:skew-pseudo-levelspacing}}
  \subfigure[Peaks in the form factor appear at
  $k\,\nicefrac{M}{N}=\nicefrac{k}{600}$, with $k\in\N$. The average
  for $\tau>1$ gives a mean degeneracy of~$13$ for the eigenangles.]
  {\includegraphics[width=0.49\textwidth]
    {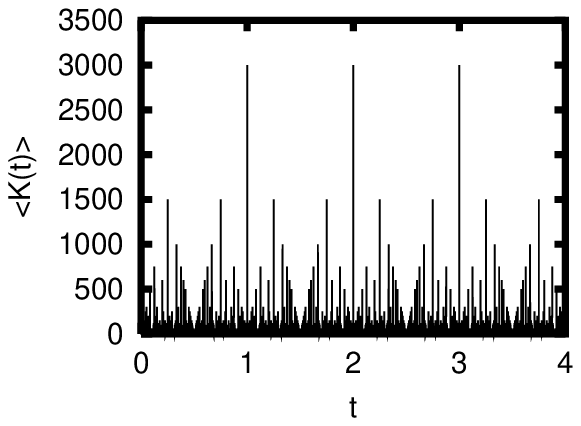}\label{fig:skew-pseudo-formfactor}}
  \caption{Spectral statistics of an almost integrable skew map with 
    $N=3000$ and $\beta=\nicefrac{3}{5}$, \ie $D=600$ and
    $M=5$.}
  \label{fig:skew-pseudo-statistics}
\end{figure}

The spectral statistics of the skew map depends sensitively on the
value of~$D$, which according to~\eqref{eq:s-qm-symmetrie-position}
determines the number of quantum symmetries.  In~\cite{BacHaa99} it
was shown that for $\beta\notin\Q$ subsequences of quantisations with
fixed~$D$ lead to different limit distributions as $N\to\infty$ for
various spectral statistics. In~\cite[corrolary 5.2]{MarRud00} it was
shown that the multiplicity of eigenphases of $\op{U}(S_\beta)$ is
bounded from above by $c(\epsilon) D^{\nicefrac{1}{2}+\epsilon}$ for
each $\epsilon>0$, where $c(\epsilon)$ does not depend on~$D$, \ie
for larger values of $D$ one can obtain more degenerate spectra.  Whereas
for $\beta\in\Q$ we have $D=\nicefrac{N}{M}$, \ie~$D$ grows linearly
with~$N$, for $\beta\notin\Q$ it can be shown~\cite{BacHaa99} that as
$N\to\infty$ there exist subsequences with fixed~$D$ for any finite
value of~$D$. For example there is always a subsequence with $D=1$. If
$\beta\notin\Q$ is badly approximable, $D$ is bounded from above by
$\tilde{c}(\epsilon) N^{\nicefrac{1}{2}+\epsilon}$ for each
$\epsilon>0$~\cite[lemma 4.2]{MarRud00}.  Thus, in general, for
rational skew maps we obtain much larger values of $D$, which
typically lead to higher degeneracies.  This is nicely illustrated in
figure~\ref{fig:skew-pseudo-levelspacing}, where $I(0)$ is larger than
in figure~\ref{fig:skew-ergodic-levelspacing}.  Roughly speaking,
quantisations of skew maps with rational~$\beta$ are even more
degenerate than those with~$\beta$ irrational.

We remark that one directly obtains an analytic expression for the
non-generic form factor in figure~\ref{fig:skew-pseudo-formfactor}
from the trace formula for the skew map, cf.
appendix~\ref{app:skew-trace}. Moreover, the mean of the form factor
for $\tau\gg 1$ gives the average of the degeneracies, which now
is~$13$ in contrast to~$\nicefrac{5}{3}$ for the ergodic example in
the previous section.

\begin{figure}
  \subfigure[The level spacing distribution shows
  intermediate statistics.]  {\includegraphics[width=0.49\textwidth]
    {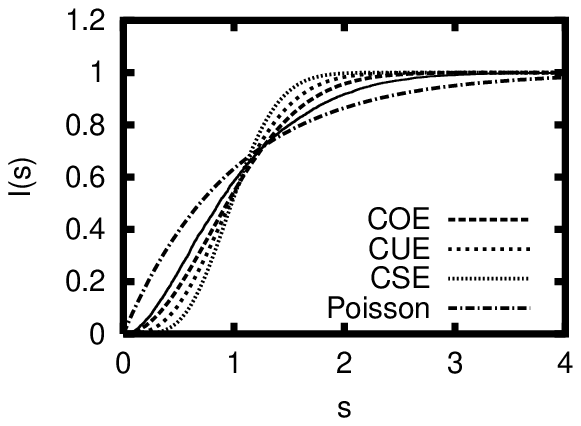}
    \label{fig:skew-spin-pseudo-levelspacing}}
  \subfigure[The smoothed form factor also shows intermediate
  statistics. For $\tau=0$ one observes a value between~$0$ and~$1$.]          
  {\includegraphics[width=0.49\textwidth]
    {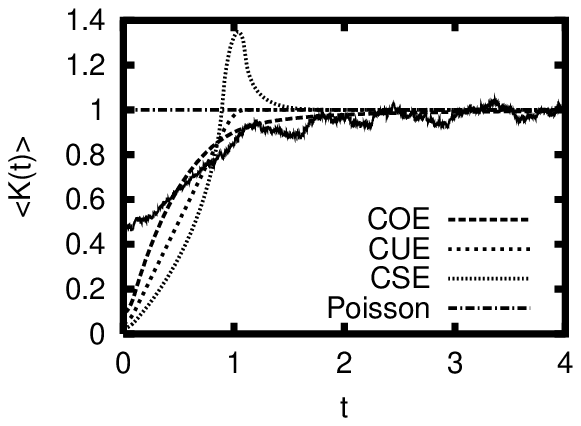}
    \label{fig:skew-spin-pseudo-formfactor}}
  \caption{Spectral statistics of an almost integrable skew map 
    with spin~$\nicefrac{1}{2}$. We have chosen the same parameters as
    in figure~\ref{fig:skew-pseudo-statistics}. Therefore we have
    $N=3000$, $\beta=\nicefrac{3}{5}$, $D=600$ and $M=5$.}
  \label{fig:skew-spin-pseudo-statistics}
\end{figure}

When now coupling a spin~$\nicefrac{1}{2}$ to the map as in the
previous cases, we again expect that this 
procedure lifts the degeneracies of
periodic orbits in the trace formula and the degeneracies of quantum
eigenphases and thus restores generic behaviour. The resulting
level spacing distribution $I(s)$ and the averaged form factor $K(\tau)$
are shown in figure~\ref{fig:skew-spin-pseudo-statistics}.  The
non-generic features in figure~\ref{fig:skew-pseudo-statistics},
caused by the high degeneracies, have vanished and we obtain an
level spacing distribution between COE and Poisson showing
level repulsion on short scales but for larger~$s$ staying below the
RMT curves. The form factor starts with a non-zero value at $\tau=0$
and then approaches~$1$.

\begin{wrapfigure}[16]{r}{0.49\textwidth}
  \includegraphics[width=0.49\textwidth]
  {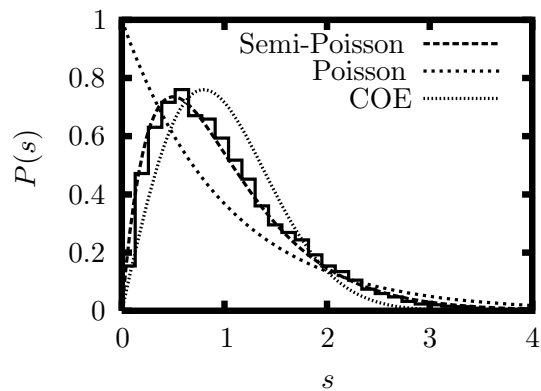}
    \caption{Level spacing density of the almost integrable 
      skew map with spin in comparison to $P_{\rm SP}(s)$.}
    \label{fig:level-spacing-pseudo}
\end{wrapfigure}
\begin{figure}[!t]
  \subfigure[The level spacing distribution varies
  with $N$. We get fluctuating about the Semi-Poissonian
  distribution.]  {\includegraphics[width=0.49\textwidth]
    {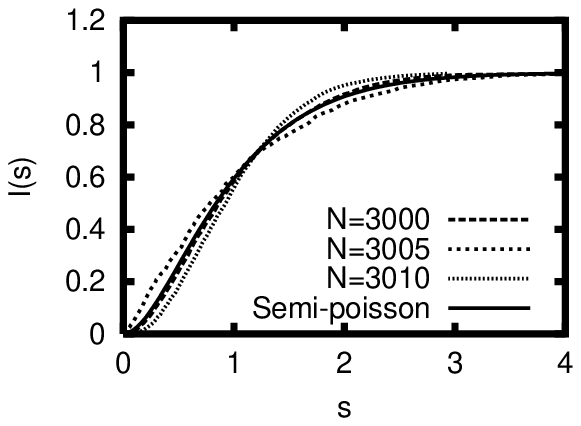}
    \label{fig:skew-spin-pseudo-levelspacing-compare}}
  \subfigure[The form factor also varies and even the value for
  $\tau=0$ changes slightly with~$N$.]
  {\includegraphics[width=0.49\textwidth]
    {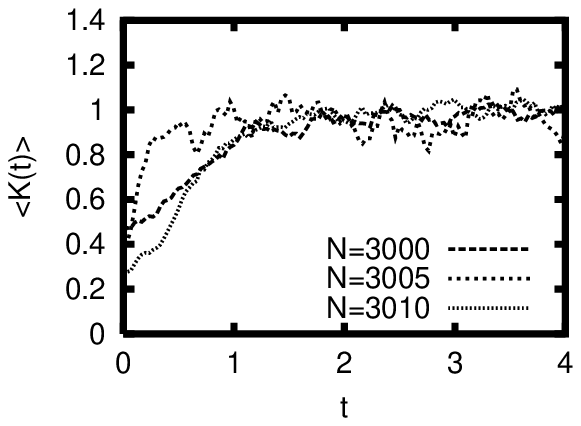}
    \label{fig:skew-spin-pseudo-formfactor-compare}}
  \caption{Comparison of spectral statistics for an almost 
    integrable skew map at different values of~$N$. We show the
    level spacing distribution and the smoothed form
    factor where we averaged over $\Delta\tau = 0.2$.}
  \label{fig:skew-spin-pseudo-statistics-compare}
\end{figure}

The intermediate level spacing distribution of almost integrable -- or
pseudointegrable -- systems, cf., \eg,
\cite{RicBer81,ShuShiSebSteStoZyc94}, is often associated with
so-called Semi-Poisson statistics \cite{BogGerSch99,BogGerSch01} which
yields
\begin{equation}
  \begin{split} 
  P_{\rm SP}(s) &= 4s\,\ue^{-2s} \quad  \text{and} \\
  I_{\rm SP}(s) &= 1 - (2s+1)\ue^{-2s}
 \ .
 \end{split} 
\end{equation}
At a first glance this appears to be consistent with
figure~\ref{fig:level-spacing-pseudo}. However, when looking more
carefully, in figure~\ref{fig:skew-spin-pseudo-statistics-compare} we
still observe fluctuations about $I_{\rm SP}(s)$ when only slightly
changing~$N$. The value of $K(0)$ also changes and, thus, from our
data we cannot judge whether the statistics will converge to a limit
distribution as $N\to\infty$. However, considering our previous
discussions it is not clear whether one should expect convergence. We
have argued that in order to observe generic statistics one has to
break the degeneracies of the quantum skew map. Since these
degeneracies grow with~$D$ this can be nicely achieved for irrational
skew maps where we obtain CUE statistics.  For rational skew maps, on
the other hand, the quickly growing~$D$ might prevent us from
restoring genericity and we might only observe a tendency towards
Semi-Poissonian statistics.

However, based on the semiclassical trace
formula~\eqref{eq:app-trace-skew-spin-7} for the skew map with spin we
can give a semi-heuristic argument yielding the value to which $K(0)$
should converge in the semiclassical limit.  If we
insert~\eqref{eq:app-trace-skew-spin-7} in the definition of the form
factor,
\begin{equation}
  K(\tau) = \frac{1}{2N} \left| \Tr \Uspin^l(S_\beta) \right|^2 - 2N 
  \delta_{0l}\ , \quad \tau = \frac{l}{2N} \ ,
\end{equation}
and perform the diagonal approximation~\cite{HanOzo84,Ber85}, \ie in
the double sum over families of periodic orbits, labelled by~$\nu$, we
keep only terms with like actions, we obtain
\begin{equation}
\label{eq:Kdiag}
  K(\tau) \sim \frac{1}{2} \sum_{k\in\Z} \delta_{l,kM} \, \frac{1}{2kM} 
  \sum_{\nu\in\Z_{2k}} M^2 \left[ 
  \big\langle \tr d_l(q,p) \big\rangle_{\E_{p^{(\nu)}}(\beta)} 
  \right]^2 \ ,
\end{equation}
where
\begin{equation}
  \big\langle f(q,p) \big\rangle_{\E_{p^{(\nu)}}(\beta)} 
  \defin \frac{1}{M} \sum_{j\in\Z_M} \int_0^1 
     f\left(q,\tilde{p}^{(\nu)}+\frac{j}{M}\right) \, \ud q
\end{equation}
denotes the average over the invariant manifold
$\E_{p^{(\nu)}}(\beta)$. Due to the anti-symplectic
symmetries~\eqref{eq:s-anti-symplectic-symmetry} there are additional
degeneracies in the actions. However, assuming that the corresponding
spin weights $\tr d_l$ are sufficiently uncorrelated we do not take
correlations between these terms into account. Thus there appears no
additional factor of~$4$ in~\eqref{eq:Kdiag}. In order to determine
$K(0)$ we average $K(\tau)$ over an interval $(0,\tau^*)$ and
simultaneously perform the limit $N\to\infty$ and $\tau^* \to 0$ such
that $\tau^* N \to \infty$. Since then the dominating contributions
come from large~$l$ we are interested in the behaviour
of~\eqref{eq:Kdiag} as $l\to\infty$. In this limit we can replace
the sum over~$\nu$ by an integral obtaining
\begin{equation}
  K(\tau) \sim \frac{1}{2} \sum_{k\in\Z} \delta_{l,kM} \, M \, \bar{a}_l
\end{equation}
with the combined average
\begin{equation}
  \bar{a}_l \defin M \int_0^{\nicefrac{1}{M}} \left[ 
  \big\langle \tr d_l(q,p) \big\rangle_{\E_{p^\prime}(\beta)} 
  \right]^2 \, \ud p^\prime \, .
\end{equation}
Performing the $\tau$-average or $l$-average, respectively, as
described above, yields
\begin{equation}
  K(0) \sim \frac{1}{2} \lim_{K\to\infty} \frac{1}{K} 
  \sum_{k=1}^K \bar{a}{_{kM}} \ , 
\end{equation}
and, since due to $|\tr d_l| \leq 2$ we have $0 \leq \bar{a}_l \leq
4$, the value of $K(0)$ is restricted to the interval $[0,2]$.

We will now give a heuristic argument why it is reasonable to expect
$K(0)\sim\nicefrac{1}{2}$. To this end consider $\big[ \left\langle
  \tr d_l(q,p) \right\rangle _{\E_{p^\prime}(\beta)} \big]^2$.  Since
the skew map~$S_\beta$ restricted to $\E_{p^\prime}(\beta)$ is ergodic
for almost all $p^\prime$ we can replace $\left\langle \ \cdot \ 
\right\rangle_{\E_{p^\prime}(\beta)}$ by some time average
$\overline{\ \cdot \ }^t$ along an orbit with initial condition
$(q_0,p_0) \in \E_{p^\prime}(\beta)$. We now assume that for large~$l$
the spin weights $\tr d_l(S_\beta^n(q_0,p_0))$ are uncorrelated, which
for the magnetic fields~\eqref{eq:B-Felder} is a reasonable
assumption. Therefore, we have $\big(\overline{\tr d_l}^t\big)^2 =
\overline{(\tr d_l)^2}^t$, which again due to the ergodicity of
$S_\beta{}$ on ${\E_{p^\prime}(\beta)}$, implies
\begin{equation}
  \bar{a}_l = M \int_0^{\nicefrac{1}{M}} \left\langle (\tr d_l(q,p))^2 
  \right\rangle_{\E_{p^\prime}(\beta)} \, \ud p^\prime \ .
\end{equation}
If we also assume that the skew product $Y^l:(q,p,g) \mapsto
(S_\beta^l(q,p),d_l(q,p) g)$, cf.~\eqref{eq:skewproduct}, is ergodic
on $\E_{p^\prime}(\beta) \times \SU(2)$ for almost all~$p^\prime$
(which, again referring to the fields~\eqref{eq:B-Felder}, is a good
assumption) we obtain, cf.~\cite{BolKep99b,KepMarMez01},
\begin{equation}
  \lim_{K\to\infty} \frac{1}{K} \sum_{k=1}^K \bar{a}{_{kM}}
  = \lim_{L\to\infty} \frac{1}{L} \sum_{l=1}^L \bar{a}{_{l}}
  = \int_{\SU(2)} (\tr g)^2 \, \ud\mu_H(g) = 1 \ , 
\end{equation}
where~$\mu_H$ denotes the normalised Haar measure on $\SU(2)$. Thus we
find $K(0) \sim \nicefrac{1}{2}$, which is in agreement with the
Semi-Poisson distribution~\cite{BogGerSch99,BogGerSch01}. However one
should note that both in the light of our previous discussion about
the slow numerical convergence and considering the various assumptions
made in the last paragraph this should not be viewed as a derivation but
rather as a semiclassical illustration of our numerical findings.  A
final decision about whether the quantised skew map with spin shows
Semi-Poissonian statistics requires further investigations, both
numerical and semiclassical.


\rcsInfo $Id: conclusions.tex,v 2.1 2001/06/26 12:02:59 guest Exp guest $
\section{Conclusions}
\label{sec:conclusions}

We have investigated quantised parabolic maps on the torus which have
highly degenerate spectra showing non-generic spectral statistics.  An
outstanding property of their classical limit is the high
degeneracy of periodic orbits occuring in continuous families. For
ergodic maps we have shown numerically that both the classical and
quantum degeneracies can be lifted by coupling the map to
spin~$\nicefrac{1}{2}$, thus leading to generic spectral statistics
following RMT predictions. Our numerical findings are supported by a
semiclassical analysis. Hence we have given evidence that in generic
situations ergodicity might be a sufficiently strong condition in order to
observe RMT statistics, sheding new light on the weakest possible
requirements for the BGS conjecture.

Moreover, we have investigated an almost integrable map which also has
a highly degenerate spectrum. Again by coupling the map to a
spin~$\nicefrac{1}{2}$ we numerically observed a transition to
intermediate statistics. We have also given a preliminary
semiclassical discussion on whether this situation can be described by
the so-called Semi-Poisson statistics. However, a final decision on
this particular issue is beyond the scope of the present paper.


\subsection*{Acknowledgement}

We would like to thank Jens Bolte, Roman Schubert and Jan Wiersig
for helpful discussion.
We also acknowledge fincancial support by the 
Deutsche Forschungsgemeinschaft (DFG)
under contracts no. Ste 241/9-2 and Ste 241/10-1.

\begin{appendix}
  \rcsInfo $Id: app.egorov.tex,v 3.1 2001/06/26 12:10:43 guest Exp guest $
\section{Egorov theorem for the Kronecker map}
\label{app:egorov}

For the sake of completeness we give a proof of the Egorov property
\eqref{eq:k-egorov} for the quantised Kronecker map,
cf.~\cite{MarRud00}.  To this end first consider the Weyl quantisation
of a time evolved observable $f(q,p)$,
\begin{equation}
\begin{split}
  \Op (f\circ\Kron^t) &= \Op \left( \sum_{n,m\in\Z} f_{nm} \,
    \ue^{2\pi\ui t(m(q+\alpha)-n(p+\beta))} \right) \\
  &= \sum_{n,m\in\Z} f_{nm} \, \ue^{2\pi\ui t(m\alpha-n\beta)} \,
  \op{T}\left(\frac{n}{N},\frac{m}{N}\right) \ ,
\end{split}
\end{equation}
with $t\in\Z$.  On the other hand, by virtue of the commutation
relations~\eqref{eq:weyl-heisenberg}, for the quantum time evolution
we have
\begin{equation}
  \op{U}^{-t}(\Kron) \, \Op(f) \, \op{U}^t(\Kron)
  = \sum_{n,m\in\Z} f_{nm} \, \ue^{2\pi\ui t \frac{ma-nb}{N}} \,
    \op{T}\left(\frac{n}{N},\frac{m}{N}\right) \ .
\end{equation}
Thus, the discrepancy between classical and quantum mechanical time
evolution is given by
\begin{equation}
\begin{split} 
  \Delta_f(t) &\defin
  \norm{\op{U}^{-t}(\Kron)\Op(f)\op{U}^t(\Kron)-\Op(f\circ \Kron^t)
    }_{\hilbert} \\
  &= \norm{ \sum_{n,m\in\Z} f_{nm} \left( \ue^{2\pi\ui t
        \frac{ma-nb}{N}} -\ue^{2\pi\ui t(m\alpha-n\beta)}
    \right) \op{T}\left(\frac{n}{N},\frac{m}{N}\right) }_{\hilbert} \\
  &\leq \sum_{n,m\in\Z} \abs{f_{nm}} \, \left| \ue^{2\pi\ui t
      \frac{ma-nb}{N}} -\ue^{2\pi\ui t(m\alpha-n\beta)} \right| \,
  \norm{ \op{T}\left(\frac{n}{N},\frac{m}{N}\right) }_{\hilbert}
\end{split}
\end{equation}
With $\norm{ \op{T}\left(\frac{n}{N},\frac{m}{N}\right) }_{\hilbert} =
1$, and an application of
\begin{equation}
  \left| \ue^{\ui zx} - \ue^{\ui zy} \right| \leq |z| |x-y| \ , \quad
  x,y,z \in \R \ ,
\end{equation}
we have
\begin{equation}
\begin{split} 
  \Delta_f(t) 
  \leq \ 
  & 2 \pi |t|  
  \sum_{n,m\in\Z}\left( \abs{\frac{a}{N}-\alpha } \abs{n}
    + \abs{\frac{b}{N}-\beta} \abs{n} \right) \abs{f_{nm}} \\
  \underset{\eqref{eq:k-approximation}}{\leq} 
  &\fracN \, \pi |t| 
  \sum_{n,m\in\Z} \big(\abs{n}+\abs{m}\big) \abs{f_{nm}} \ ,
\end{split} 
\end{equation}
which concludes the proof, since $f\in C^\infty(\TT)$ implies that 
$f_{nm}$ decays exponentially as $n,m\to\infty$,

 
  \rcsInfo $Id: app.skew-trace.tex,v 3.2 2001/06/26 12:13:32 guest Exp guest $
\section{Trace formula for skew maps}
\label{app:skew-trace}
Before calculating the trace formula for the skew map we will
summarise the structure of periodic orbits~$\gamma$ for a skew map
with a rational translation~$\nicefrac{b}{N}$. From the definition of
the map we get the condition for periodic orbits
\begin{equation}
  \label{eq:app-po}
  \begin{pmatrix}
    p_l \\ q_l
  \end{pmatrix}
  =
  \begin{pmatrix}
    p_0+l\fracN[b] \\ q_0+2lp_0+l(l-1)\fracN[b]
  \end{pmatrix}
  \stackrel{!}{=}
  \begin{pmatrix}
    p_0 \\ q_0
  \end{pmatrix}
  \ \pmod 1
  \ .
\end{equation}
The period~$l_{\gamma}$ is determined by the condition in momentum.
With
\begin{equation}
  \label{eq:app-abbreviation}
  D\defin\gcd(b,N)
  \quad\text{and}\quad
  M\defin\frac{N}{D}
\end{equation}
we get
\begin{equation}
  \label{eq:app-period}
  l_{\gamma}=kM  
  \ ,\quad\text{with}\quad 
  k\in\N
  \ .
\end{equation}
The other equation gives the momenta of the periodic orbits,
\begin{equation}
  \label{eq:app-momenta-po}
  p_0=\frac{\nu}{2kM}
  \ ,\quad\text{with}\quad 
  \nu\in\ZN[2kM]
  \ . 
\quad 
\end{equation}
There is no restriction on the initial position~$q_0$ and thus the
periodic orbits occur in one-parameter families.

After this preliminary remark we now turn to the calculation of the
trace formula. Starting point is the one-step
propagator~\eqref{eq:s-propagator-mixed} which in position
representation reads
\begin{equation}
  \label{eq:app-propagator-position}
  \braopket{Q}{\op{U}^{\vphantom{l}}(\Skew)}{Q'}
  =\fracN\sum_{P\in\ZN}
  \exp\left[\zpi N \left(G_{\nicefrac{b}{N}}
      \left(\fracN[Q],\fracN[P]\right)
      -\fracN[Q']\fracN[P]\right)
  \right]
  \ .
\end{equation}
One easily verifies by induction that the $l$-step propagator is given
by
\begin{equation}
  \label{eq:app-l-step-propagator}
  \braopket{Q}{\op{U}^l(\Skew)}{Q'}
  =\fracN\sum_{P\in\ZN}
  \exp\left[\zpi N \left(G^l_{\nicefrac{b}{N}}
      \left(\fracN[Q],\fracN[P]\right)
      -\fracN[Q']\fracN[P]\right)\right]
  \ ,
\end{equation}
where
\begin{equation}
  \label{eq:app-generating-1}
  G^l_{\nicefrac{b}{N}}\left(q_l,p_0\right)
  \defin q_l\left(p_0+l\fracN[b]\right)
  -\sum_{\lambda=0}^{l-1}\left(p_0+\lambda\fracN[b]\right)^2
\end{equation}
denotes the generating function for the $l$-step skew map from
$(p_0,q_0)$ to $(p_l,q_l)$, \ie
\begin{equation}
  \label{eq:app-generating-2}
  p_{l}
  =\frac{\partial G^l_{\nicefrac{b}{N}}\left(q_{l},p_0\right)}
  {\partial q_{l}}
  \quad \text{and} \quad 
  q_0
  =\frac{\partial G^l_{\nicefrac{b}{N}}\left(q_{l},p_0\right)}
  {\partial p_0}
  \ .
\end{equation}
Applying the Poisson summation formula to the
propagator~\eqref{eq:app-l-step-propagator} yields
\begin{equation}
  \label{eq:app-l-step-propagator-poisson}
  \braopket{Q}{\op{U}^l(\Skew)}{Q'}
  =\sum_{n\in\Z}\int_0^1
  \exp\left[\zpi N\left(G^l_{\nicefrac{b}{N}}\left(\fracN[Q],p\right)
      -\fracN[Q']p+np\right)\right]\intd p
  \ .
\end{equation}
We can split the sum over~$n$ by writing $n=2l\mu+\nu$, $\mu\in\Z$,
$\nu \in \Z_{2l}$ and thus we get \setlength{\mathindent}{0cm}
\begin{equation}
  \label{eq:app-l-step-propagator-split}
  \braopket{Q}{\op{U}^l(\Skew)}{Q'}
  =\sum_{\nu\in\ZN[2l]}\sum_{\mu\in\Z}\int_0^1
  \exp\left[\zpi N\left(G^l_{\nicefrac{b}{N}}\left(\fracN[Q],p\right)
      -\fracN[Q']p+(2l\mu+\nu)p\right)\right]\intd p
  \ .
\end{equation}
With the substitution $p'=p+\mu$ we can rearrange the integrals such
that
\begin{equation}
  \label{eq:app-l-step-propagator-substitute-2}
    \braopket{Q}{\op{U}^l(\Skew)}{Q'}
    =\sum_{\nu\in\ZN[2l]}\int_{-\infty}^{\infty}
    \exp\left[\zpi N\left(G^l_{\nicefrac{b}{N}}\left(\fracN[Q],p'\right)
        -\left(\fracN[Q']-\nu\right)p'\right)\right]\intd p'
    \ .
\end{equation}
The (Gaussian) integral can be calculated yielding
\begin{equation}
  \label{eq:app-l-step-propagator-action}
  \braopket{Q}{\op{U}^l(\Skew)}{Q'}
  =\sqrt{\frac{-\ui}{2lN}}\sum_{\nu\in\ZN[2l]}
  \exp\left[\zpi N S^l_{\nicefrac{b}{N}}\left(\fracN[Q]
      +\nu,\fracN[Q']\right)\right]
  \ ,
\end{equation}
where we have taken the freedom to change~$\nu$ to~$-\nu$. We denote
by $S^l_{\nicefrac{b}{N}}(q,q')$ the Legendre transform of
$G^l_{\nicefrac{b}{N}}\left(q_l,p_0\right)$ with respect to~$p_0$,
\begin{equation}
  \label{eq:app-l-step-action}
  S^l_{\nicefrac{b}{N}}(q_l,q_0)
  =\frac{1}{2l}\left(q_l-q_0\right)^2
  +\frac{l+1}{2}\fracN[b]q_l
  +\frac{l-1}{2}\fracN[b]q_0
  -\frac{1}{4}\sum_{\lambda=0}^l \lambda(\lambda-1)
  \ ,
\end{equation}
with
\begin{equation}
  \label{eq:app-l-step-generating}
  p_{l}
  =\frac{\partial S^l_{\nicefrac{b}{N}}\left(q_{l},q_0\right)}
  {\partial q_{l}}
  \quad \text{and} \quad 
  p_0
  =-\frac{\partial S^l_{\nicefrac{b}{N}}\left(q_{l},q_0\right)}
  {\partial q_0}
  \ .
\end{equation}
Therefore, the initial momentum $p^{(\nu)}$ of an orbit from
~$\nicefrac{Q'}{N}$ to $\nicefrac{Q}{N}+\nu$ with length~$l$ is given
by
\begin{equation}
  \label{eq:app-momentum-orbit}
  p^{(\nu)}
  =\left(\frac{Q-Q'}{2lN}+
    \frac{\nu}{2l}+\frac{l-1}{2}\fracN[b]\right)\pmod 1
  \ ,
\end{equation}
where~$\nu$ labels the~$2l$ different orbits from~$\nicefrac{Q'}{N}$
to~$\nicefrac{Q}{N}$.  Hence the propagator in position representation
is given by the sum over orbits from~$\nicefrac{Q'}{N}$
to~$\nicefrac{Q}{N}$.  We can now calculate the trace of the
propagator~\eqref{eq:app-l-step-propagator-substitute-2},
\begin{equation}
  \label{eq:app-trace-1}
  \Tr \op{U}^l(\Skew)
  =\sum_{Q\in\ZN}\braopket{Q}{U^l(\Skew)}{Q}
  \ .
\end{equation}
By using eq.~\eqref{eq:app-l-step-action} we get
\begin{equation}
  \label{eq:app-trace-2}
    \Tr \op{U}^l(\Skew)
    =\sqrt{\frac{N}{2\ui l}}\sum_{k\in\Z}\delta_{l,kM}\sum_{\nu\in\ZN[2l]}
    \exp\left[\zpi N \fancyS^l_\nu\right]
    \ ,
\end{equation}
with
\begin{equation}
  \label{eq:app-trace-3}
   \fancyS^l_\nu
   \defin S^l_{\nicefrac{b}{N}}\left(q+\nu,q\right)-l\fracN[b]q
   \ .
\end{equation}
One easily verifies that the right-hand side of~\eqref{eq:app-trace-3}
does not depend on the position~$q$. Notice that $l\,\nicefrac{b}{N}$
is the winding number in momentum.  Thus for periodic orbits with
fixed momentum $p$ we have the same action~$\fancyS^l_\nu$ for all
$q\in\Sone$.

It remains to show that the structure of the trace formula remains the
same when the coupling to the spin degrees of freedom is introduced.
To this end we have to calculate (cf.
section~\ref{sec:trace-formula-with-spin} and~\cite{KepMarMez01})
\begin{equation}
  \label{eq:app-trace-skew-spin-1}
  \Tr \Uspin^l(S_\beta) = \Tr \left( \op{U}(S_\beta) \Op(f) \right) 
\end{equation}
with $f = \tr d_l$. With~\eqref{eq:quantum-observables}
and~\eqref{eq:app-l-step-propagator-action} we have
\begin{equation}
  \label{eq:app-trace-skew-spin-2}
\begin{split} 
  \Tr \Uspin^l(S_\beta) = \sum_{Q\in\Z_N} & \frac{1}{\sqrt{2\ui lN}}
  \sum_{\nu\in\Z_{2l}} \sum_{m,n \in \Z} \exp\left( 2\pi\ui N
    S_{\nicefrac{b}{N}}^l\left( \frac{Q}{N} - \nu, \frac{Q-m}{N}
    \right) \right) \\
  & \times \ f_{nm} \exp\left(\ui\pi\frac{nm}{N}+2\pi\ui\frac{Qn}{N}
  \right) \ .
\end{split}  
\end{equation}
For $f\in C^\infty(\T^2)$ we may expand the action as $N\to\infty$
(cf.~\cite{KepMarMez01}),
\begin{equation}
  \label{eq:app-trace-skew-spin-3}
\begin{split}   
  S_{\nicefrac{b}{N}}^l\left( \frac{Q}{N} - \nu, \frac{Q-m}{N} \right)
  &\sim S_{\nicefrac{b}{N}}^l\left( \frac{Q}{N} - \nu, \frac{Q}{N}
  \right) - \frac{\partial S_{\nicefrac{b}{N}}^l}{\partial q_0}
  \left( \frac{Q}{N} - \nu, \frac{Q}{N} \right) \frac{m}{N} \\
  &= \fancyS^l_\nu + l \frac{b}{N} \frac{Q}{N} + \frac{m}{N}
  p^{(\nu)}\ .
\end{split}
\end{equation}
Now by the same considerations as in section~\ref{sec:trace-formula}
for~$l$ fixed and~$N$ large the sum over the positions~$Q$ becomes
\begin{equation}
  \label{eq:app-trace-skew-spin-4}
  \sum_{Q\in\Z_n} \exp\left(\frac{2\pi\ui}{N}(lb+n)Q\right) 
  \sim \delta_{n0} \sum_{k\in\Z} N \, \delta_{l,kM} \ ,
\end{equation}
\ie it selects the periods~$kM$ of periodic orbits of the skew map.
Thus, the trace of the propagator reads
\begin{equation}
  \label{eq:app-trace-skew-spin-5}
  \Tr \Uspin^l(S_\beta) \sim \sum_{k\in\Z} \delta_{k,lM}  
  \, \sqrt{\frac{N}{2\ui l}}
  \sum_{\nu\in\Z_{2l}} \ue^{2\pi\ui N\fancyS^l_\nu}
  \sum_{m\in\Z} f_{0m} \ue^{2\pi\ui mp^{(\nu)}} \, .
\end{equation}
By~\eqref{eq:classical-observables} the sum over~$m$ is given by
$\int_0^1 f(p^{(\nu)},q) \intd q$, \ie resubstituting $f=\tr d_l$ we
obtain
\begin{equation} 
 \label{eq:app-trace-skew-spin-6}
  \Tr \Uspin^l(S_\beta) 
  \sim \sum_{k\in\Z} \delta_{k,lM} \, 
  \sqrt{\frac{N}{2\ui l}} \sum_{\nu\in\Z_{2l}} 
       \ue^{2\pi\ui N\fancyS^l_\nu}
       \int_0^1 \tr d_l\!\left( q, p^{(\nu)} \right) \intd q \ .
\end{equation}
Observing that $\fancyS^l_\nu = \fancyS^l_{\nu^\prime}$ if
$p^{(\nu^\prime)} \in \E_{p^{(\nu)}}(\nicefrac{b}{N})$,
cf.~\eqref{eq:disjoint-circles}, which can be checked by direct
computation, instead of summing over the momenta~$p^{(\nu)}$ of
periodic points we can sum over the invariant manifolds
$\E_{p_0}(\nicefrac{b}{N})$ and their disconnected components.
Finally we have derived the trace formula
\begin{equation} 
 \label{eq:app-trace-skew-spin-7}
  \Tr \Uspin^l(S_\beta) 
  \sim \sum_{k\in\Z} \delta_{l,kM} \, 
  \sqrt{\frac{N}{2\ui l}} \sum_{\nu\in\Z_{2k}} 
       \ue^{2\pi\ui N\fancyS^l_\nu} \sum_{j\in\Z_M}
       \int_0^1 
       \tr d_l\!\left( q, \tilde{p}^{(\nu)}+\frac{j}{M} \right) 
       \intd q 
  \ ,
\end{equation}
where, $\tilde{p}^{(\nu)} \in \E_{p^{(\nu)}}(\nicefrac{b}{N})$ can be
chosen such that $0 \leq \tilde{p}^{(\nu)} < \nicefrac{1}{M}$.  Now
the right-hand side of~\eqref{eq:app-trace-skew-spin-7} has to be
interpreted as follows: We have obtained a sum over all periodic
orbits of the (approximating) classical skew map. The periodic orbits
appear in families labelled by $\nu$ and the spin weights $\tr d_l$
are integrated over the corresponding invariant manifold
$\E_{p^{(\nu)}}(\nicefrac{b}{N})$.


\end{appendix}

\bibliographystyle{my_unsrt.bst}
\bibliography{referenzen.bib}

\end{document}